\documentclass{article}
\usepackage{amsmath,amssymb,graphicx,fullpage}

\usepackage{color}
\usepackage{bbm}

\newcommand{\cI}{\mathcal{I}}
\newcommand{\cJ}{\mathcal{J}}
\newcommand{\Omegax}{\Omega_{\mathbf{x}}}
\newcommand{\Omegak}{\Omega_{\mathbf{k}}}
\newcommand{\R}{\mathbb{R}}
\newcommand{\tenp}{\bar{\mathbf{p}}}
\newcommand{\tenv}{\bar{\mathbf{v}}}

\newcommand{\modify}{\color{black}}
\let\oldbibliography\thebibliography
\renewcommand{\thebibliography}[1]{%
  \oldbibliography{#1}%
  \setlength{\itemsep}{0pt}%
}

\usepackage[colorlinks=true,
            linkcolor=red,
            urlcolor=black,
            citecolor=blue]{hyperref}

\usepackage{tabularx}
\usepackage{multirow}

\begin{document}

\title{Tensor methods for parameter estimation and bifurcation 
analysis of stochastic reaction networks}

\author{Shuohao Liao\thanks{Mathematical Institute, University of Oxford, 
Radcliffe Observatory Quarter, Woodstock Road, Oxford OX2 6GG, 
United Kingdom; e-mails: liao@maths.ox.ac.uk; erban@maths.ox.ac.uk},
\and
Tom\'{a}\v{s} Vejchodsk\'{y}\thanks{Institute of Mathematics, 
Czech Academy of Sciences, Zitna 25, 115 67 Praha 1, Czech Republic; 
e-mail vejchod@math.cas.cz},
\and
Radek Erban$^*$}

\maketitle

\begin{abstract}
\noindent
Stochastic modelling of gene regulatory networks provides 
an indispensable tool for understanding how random events at 
the molecular level influence cellular functions. A common
challenge of stochastic models is to calibrate a large number 
of model parameters against the experimental data. Another 
difficulty is to study how the behaviour of a stochastic model 
depends on its parameters, i.e. whether a change in model 
parameters can lead to a significant qualitative change in 
model behaviour (bifurcation). In this paper, tensor-structured 
parametric analysis (TPA) is developed to address these 
computational challenges. It is based on recently proposed 
low-parametric tensor-structured representations of classical 
matrices and vectors. This approach enables simultaneous 
computation of the model properties for all parameter values 
within a parameter space. The TPA is illustrated by 
studying the parameter estimation, robustness, sensitivity 
and bifurcation structure in stochastic models of biochemical 
networks. A Matlab implementation of the TPA is available 
at \texttt{http://www.stobifan.org}.
\end{abstract}

\section{Introduction}

Many cellular processes are influenced by stochastic 
fluctuations at the molecular level 
which are often modelled using stochastic simulation algorithms 
for chemical reaction 
networks~\cite{ozbudak2002regulation,swain2002intrinsic}.  For example, 
cell metabolism, signal transduction 
and cell cycle can be described by network structures of functionally 
separated modules of gene 
expression~\cite{li2010biomodels}, the so called gene regulatory 
networks (GRNs). 

Typical GRN models can have tens of variables and parameters. 
Traditionally, GRNs have been 
described using continuous deterministic models written as systems 
of ordinary differential equations (ODEs).
Several methodologies for studying parametric 
properties of  ODE systems,
such as identifiability and bifurcation, 
have been developed in the 
literature~\cite{tsai2008robust, novak2008design, % 
kitano2004biological,brophy2014principles, karlebach2008modelling}.
Recently, experimental evidence has highlighted the significance of 
intrinsic randomness in GRNs, and stochastic models have been increasingly 
used~\cite{ozbudak2002regulation,mcadams1997stochastic}.
They are usually simulated using the Gillespie stochastic 
simulation 
algorithm (SSA)~\cite{gillespie1977exact}, or its equivalent 
formulations~\cite{cao2004slow,gibson2000efficient}.
However, methods for parametric analysis of ODEs cannot be directly
applied to stochastic models.
In this paper, we present a \emph{tensor-structured parametric 
analysis} (TPA) which can be used to understand how 
molecular-level fluctuations influence the system-level behaviour of 
GRNs and its dependence on model parameters. We illustrate major 
application areas of the TPA by studying 
several biological models with increasing level of complexity. 

The parametric analysis of GRN models is computationally intensive 
because both state space
and parameter space are high-dimensional. The dimension of the 
state space, $\Omega_{\mathbf{x}}$, 
is equal to the number of reacting molecular species, denoted by $N$. 
When an algorithm, 
previously working with deterministic steady states, is extended to 
stochastic setting, its computational complexity
is typically taken to the power $N$. Moreover, the exploration of 
the parameter space, $\Omega_{\mathbf{k}}$,
introduces another multiplicative exponential complexity. Given 
a system that involves $K$
parameters, the `amount' of parameter combinations to be characterized 
scales equally with 
the volume of $\Omega_{\mathbf{k}}$, i.e. it is taken to the power 
$K$~\cite{hey2007integration}.

{\modify

The TPA framework avoids the high computational cost of working in high-dimensional $\Omega_\bold{x}$ and $\Omega_\bold{k}$. The central idea is based on generalising the concept of separation of variables to parametric probability distributions~\cite{beylkin2002numerical}. The TPA framework can be divided into 
two main steps: a tensor-structured computation and a tensor-based analysis. 
First, the steady state distributions of stochastic models are simultaneously 
computed for all possible parameter combinations within a parameter space and 
stored in a tensor format, with smaller computational and memory requirements 
than in traditional approaches. The resulting tensor data are then analyzed 
using algebraic operations with computational complexity which scales linearly 
with dimension (i.e., linearly with $N$ and $K$). 

The rest of this paper is organised as follows. In Section 2, we discuss how 
the parametric steady state probability distribution can be presented and computed 
in tensor formats. We illustrate the data storage savings using tensor-structured 
simulations of four biological systems. The stored tensor data are then used as 
the input for the tensor-based analysis presented in the subsequent sections. 
In Section 3, we show that the existing procedures for parameter inference 
for deterministic models can be directly extended to the stochastic models 
using the computed tensor data. In Section 4, a direct visualisation of  
stochastic bifurcations in a high-dimensional state space is presented. 
The TPA of the robustness of the network to extrinsic noise is illustrated
in Section 5. We conclude with a brief discussion in Section 6.

}

\section{{\modify Tensor-structured computations}}

Considering a well-mixed chemically-reacting system of $N$ 
distinct molecular species  
$X_i$, $i=1,2,\dots,N$, inside a reactor (e.g. cell) of volume  $V$, 
we denote its state vector by
$\mathbf{x} = (x_1, x_2, \ldots,x_N)^T$, where $x_i$ is the number of 
molecules of the $i$-th
chemical species $X_i$. In general, the volume $V$ can be time 
dependent (for example, in cell cycle models which explicitly take 
into account cell growth), but we will focus in this
paper on models with constant values of $V$. We assume that molecules 
interact through $M$ reaction channels
\begin{equation}
\sum_{i=1}^N \nu^-_{j,i}X_i
\overset{k_j}{\longrightarrow} 
\sum_{i=1}^N \nu^+_{j,i}X_i,
\quad j=1,2,\dots,M,
\label{reactionsystem}
\end{equation}
where $\nu^+_{j,i}$ and $\nu^-_{j,i}$ are the stoichiometric coefficients. 
The kinetic rate 
parameters, $\mathbf{k} = (k_1,k_2,\ldots, k_M)^T$, characterise the 
rate of the corresponding 
chemical reactions. We will treat $\mathbf{k}$ as auxiliary variables, 
and in other words, 
the parametric problem of (\ref{reactionsystem}) involves 
considering both 
$\mathbf{x} \in \Omega_{\mathbf{x}}$ and $\mathbf{k} \in \Omega_{\mathbf{k}}$.
In this paper, we study problems where the dimension of the parameter 
space $K$
is equal to $M$. We also consider cases where some rate constants are 
not varied
in the parameter analysis, i.e. $K < M$. In this case, notation 
$\mathbf{k}$ will be
used to denote $K$-dimensional vector of rate constants, 
$\mathbf{k} = (k_1,k_2,\ldots, k_K)^T$, 
which are considered during the TPA. The values of the 
remaining $(M-K)$ rate constants
are fixed. In principle, the TPA could also be used to study models where 
$K > M$, i.e. when we consider additional parameters (e.g., system volume $V$).

Let $p(\mathbf{x} | \mathbf{k})$ be the steady state probability 
distribution that the
state vector is $\mathbf{x}$ (if the system is observed for sufficiently 
long time)
given the parameter values $\mathbf{k}$. 
The main idea of the TPA is to split $p(\mathbf{x} | \mathbf{k})$ in 
terms of coordinates as
\begin{equation}
p(\mathbf{x} | \mathbf{k}) = 
\sum_{\ell=1}^R 
\underset{\Omega_{\mathbf{x}}}{\underbrace{f_1^\ell(x_1) \cdots f_N^\ell(x_N)}}
\ 
\underset{\Omega_{\mathbf{k}}}{\underbrace{g_1^\ell(k_1) 
\cdots g_K^\ell(k_K)}},
\label{tensorrep}
\end{equation}
where $\{ f_i^\ell(x_i) \}_{i=1,\ldots,N}$ and 
$\{ g_j^\ell(k_j) \}_{j=1,\ldots,K}$ are 
univariate functions that vary solely with a single state variable and 
parameter,
respectively. The number of summands $R$, the so 
called \emph{separation rank},
controls the accuracy of the decomposition (\ref{tensorrep}).
By increasing $R$, the separated expansion could theoretically 
achieve arbitrary accuracy.

{\modify
The value of the separation rank $R$ can be analytically computed for 
simple systems. For example, there are analytical formulas for the stationary 
distributions of first-order stochastic reaction networks~\cite{jahnke2007solving}.
They are given in the form (\ref{tensorrep}) with $R=1$. Considering
second-order stochastic reaction networks, there are no general analytical 
formulas for steady state distributions. They have to be approximated
using computational methods. The main assumption of the TPA approach 
is that the parametric steady state distribution has a sufficiently accurate
low-rank representation (\ref{tensorrep}). In this paper, we show that this 
assumption is satisfied for realistic biological systems by applying the
TPA to them and presenting computed (converged) results. The main consequence
of low-rank representation (\ref{tensorrep}) is that mathematical operations 
on the probability distribution $p(\mathbf{x}|\mathbf{k})$ in $N + K$ 
dimensions can be performed using combinations of one-dimensional 
operations, and the storage cost is bounded by $(N + K)R$.
The rank $R$ may also depend on $N + K$ and the size of the univariate 
functions in (\ref{tensorrep}). Numerical experiments have shown a linear 
growth of $R$ with respect to $N + K$ and 
a logarithmic growth with respect to the size of the univariate functions 
in the representation (\ref{tensorrep})~\cite{dolgov2012,Khoromskij2011}. }To find the representation (\ref{tensorrep}),  we solve the chemical Fokker-Planck equation (CFPE), 
as a (fully) continuous approximation to a (continuous time) discrete space 
Markov chain described by the corresponding chemical master equation 
(CME)~\cite{gillespie1996multivariate,Erban:2009:ASC}. 
Specifically, we keep all the objects in the separated form of (\ref{tensorrep}) during 
the computations, such that exponential scaling in complexity does not apply 
during any step of the TPA.

We refer to the representation (\ref{tensorrep}) as 
tensor-structured, because computations are performed 
on $p(\mathbf{x} | \mathbf{k})$ as multidimensional arrays of real 
numbers, which we call tensors~\cite{pereyra1973efficient}. 
The (canonical) tensor decomposition~\cite{carroll1970analysis}, 
as a discrete counterpart of (\ref{tensorrep}), 
then allows a multidimensional array to 
be approximated as a sum of tensor products of one-dimensional vectors. 
Within such a format, we can define standard algebraic operations similar 
to standard matrix operations such that the resulting tensor calculus 
enables efficient computation. The tensor-structured parametric steady 
state distribution (\ref{tensorrep}) is approximated 
as the eigenfunction corresponding to the smallest eigenvalue of the 
parametric Fokker-Planck operator. 
The operator is constructed in a tensor separated representation as 
a sum of tensor products of one-dimensional 
operators. The eigenfunction is computed by adaptive shifted inverse 
power method, using minimum alternating
energy method as the linear solver. 
We leave further discussion of technical 
computational details of the underlying methods to \emph{Supporting 
Information (SI) Appendix} S1.
The TPA has been implemented in MATLAB, 
and is part of the Stochastic Bifurcation Analyzer toolbox  
available at \texttt{http://www.stobifan.org}. 
The source code relies on the Tensor Train 
Toolbox~\cite{oseledets2011tensor}. 

\subsection{Applications of the TPA to biological systems}

\label{sectab1}

\begin{table}[t]
\caption{{\it Comparison of the matrix-based and 
tensor-structured methodologies.}}
\label{tabsimdetails}
{\small \vskip 2mm
\begin{tabular*}{\hsize}
{@{\extracolsep{\fill}} @{\vrule height 10pt depth0pt  width0pt} lllllllllll}
\hline
& \multicolumn{3}{|l}{Dimensionality} 
& \multicolumn{2}{|l}{Matrix-based} 
& \multicolumn{4}{|l}{Tensor-structured}\\
\noalign{\vskip-12pt}
Biochemical \\ 
\cline{2-10}
\vrule  width 0pt system 
& \multicolumn{1}{|c}{$N$} & \multicolumn1c{$K$} & \multicolumn1c{$N+K$}
&  \multicolumn{1}{|c}{$\text{Mem}_{\text{CME}}^{\dagger}$} & \multicolumn1c{$\text{Mem}_{\text{CFPE}}^{\dagger}$} 
& \multicolumn{1}{|c}{$\text{Mem}_\mathbf{A}$} & \multicolumn1c{$\text{Mem}_{p}$} & 
\multicolumn1c{$T_\mathbf{A}$[sec]} & \multicolumn1c{$T_\mathrm{tot}$[min]}\cr
%\noalign{\vskip-5pt}
\hline
Schl\"{o}gl & \multicolumn{1}{|c}{1} 
& 4& 5& \multicolumn{1}{|c}{$2.68\times 10^{13}$} 
& 
$2.74 \times 10^{11}$ & \multicolumn{1}{|c}{$4.01 \times 10^3$}
& $2.07 \times 10^5$ & $1.2$ & $30$\\
Cell cycle & \multicolumn{1}{|c}{6} 
& 1& 7 & \multicolumn{1}{|c}{$6.68 \times 10^{17}$} 
& 
$7.04 \times 10^{13}$ & \multicolumn{1}{|c}{$2.96 \times 10^4$} 
& $1.00 \times 10^7$  & $1.1$ & $6433$ \\
FitzHugh-Nagumo & \multicolumn{1}{|c}{2} 
& 4& 6& \multicolumn{1}{|c}{$6.38 \times 10^{14}$} & 
$1.75 \times 10^{13}$ & \multicolumn{1}{|c}{$7.65 \times 10^4$} 
&$4.02 \times 10^5$ & $0.7$ & $37$ \\
Reaction chain & \multicolumn{1}{|c}{20} 
& 0 & 20 & \multicolumn{1}{|c}{$1.20 \times 10^{44}$} & 
$1.53 \times 10^{54}$ & \multicolumn{1}{|c}{$9.26 \times 10^4$}
 & $7.28 \times 10^5$ & $15.6$ & $283$ \cr \hline
\end{tabular*}
}
{\modify
$^{\dagger}$ {\footnotesize Estimated as the product of the number of discrete 
states and the number of parameter values.}} 
\end{table}

We demonstrate the capabilities of the TPA framework by investigating 
four examples of stochastic reaction networks: 
a bistable switch in the 5-dimensional Schl\"{o}gl
model~\cite{schlogl1972chemical},
oscillations in the 7-dimensional cell cycle model~\cite{tyson1991modeling}, 
neurons excitability in the 6-dimensional FitzHugh-Nagumo 
system~\cite{tsai2008robust} and a 20-dimensional reaction 
chain~\cite{gillespie2002chemical} 
(see \emph{SI Appendix} S2 for more details of these models). 
Table~\ref{tabsimdetails} compares  
computational performance of the TPA with the traditional matrix-based 
methods for the computation of the parametric steady state distribution 
$p(\mathbf{x}|\mathbf{k})$.
The minimum memory requirements of solving the CME and the CFPE using 
matrix-based 
methods, $\text{Mem}_{\text{CME}}$ and $\text{Mem}_{\text{CFPE}}$, are 
estimated as 
products of numbers of discrete states times the total number of parameter 
combinations.
They vary in ranges $10^{13}$--$10^{44}$ and $10^{11}$--$10^{54}$, 
respectively, which 
are beyond the limits of the available hardware. In contrary, the TPA 
maintains affordable 
computational and memory requirements for all four problems considered,
as we show in Table~\ref{tabsimdetails}.
The major memory requirements of the TPA are $\text{Mem}_\mathbf{A}$
and $\text{Mem}_p$ to store the discretized Fokker-Planck operator
and the steady state distribution $p(\mathbf{x} | \mathbf{k})$, respectively
(see \emph{SI Appendix} S1 for detailed definitions).
Similarly, $T_\mathbf{A}$ is the computational time
to assemble the operator and $T_\mathrm{tot}$ is the total computational time.

{\modify
Table~\ref{tabsimdetails} shows that the TPA can outperform standard
matrix-based methods. It can also be less computationally intensive than
stochastic simulations in some cases. For example, the total computational 
time is around 30 minutes for the TPA to simulate $64^4$ different parameter 
combinations within the 4-dimensional parameter space of 
the Schl\"{o}gl chemical system (see Table~\ref{tabsimdetails}). 
If we wanted to compute 
the same result using the Gillespie SSA, we would have to run $64^4$ 
different stochastic simulations. If they had to be
all performed on one processor in 30 minutes, then we would only have 
$1.07 \times 10^{-4}$ second per one stochastic simulation and it would not
be possible to estimate the results with the same level of accuracy. 
In addition the TPA directly provides the steady state distribution
$p(\mathbf{x} | \mathbf{k})$, which would be computationally intensive 
to obtain by stochastic simulations (with the same level of accuracy) for
larger values of $N+K$.
}

{\modify 
\section{Parameter estimation} 
}

{\modify

Small uncertainties in the reaction rate values of stochastic reaction 
networks (\ref{reactionsystem}) are common in applications. Some model
 parameters are difficult to 
measure directly, and instead are estimated by fitting to time-course 
data. If GRNs are modelled using deterministic ODEs, there is 
a wide variety of tools available for parameter estimation. Many simple 
approaches are non-statistical~\cite{Moles2003}, and the procedure usually, 
although not necessarily~\cite{Jaqaman2006}, follows the algorithm presented 
in Table~\ref{algor1}. We tune uncertain parameters to minimize the distance
measure $d(\hat{\bold{x}}, \bold{x}^*)$, while the rules to generate 
candidate parameters $\bold{k}^*$ in step (a1) and the definition of 
distance function along with stopping criteria in step (a3) may vary 
in different methods. In optimization-based methods, $\bold{k}^*$ may follow 
the gradient on the surface of the distance function~\cite{Moles2003}. 
In statistical methods, such distance measure is 
provided in the concept of likelihood, 
$ \mathcal{L}(\bold{k}^* | \hat{\bold{x}}) 
=  
p(\hat{\bold{x}} | \bold{k}^*)$~\cite{Pedersen1995}. In Bayesian methods, 
the candidate parameters $\bold{k}^*$ are generated from some prior 
information 
regarding uncertain parameters, $\pi (\bold{k})$, and form a posterior 
distribution rather than a single point estimate~\cite{Toni2009}.

\begin{table}[t]
\caption{{\it {\modify Parameter estimation for ODEs.}}}
\label{algor1}
{\modify
\vskip 2mm
\hrule
\vskip 2mm
{\bf (a1)}  Generate a candidate parameter vector 
$\bold{k}^* \in \Omega_{\bold{k}}$.\\
{\bf (a2)}  Compute model prediction $\bold{x}^*$ using the parameter 
vector $\bold{k}^*$.\\
{\bf (a3)} Compare the simulated data $\bold{x}^*$ with the experimental 
evidence $\hat{\bold{x}}$, using distance function 
$d(\hat{\bold{x}}, \bold{x}^*) $ \\ 
\rule{0pt}{1pt} \hskip 7.5mm and tolerance 
$\varepsilon$. If $d(\hat{\bold{x}}, \bold{x}^*) < \varepsilon$, then accept $\bold{k}^*$. The tolerance $\varepsilon > 0$ is the desired level of 
agreement \\
\rule{0pt}{1pt} \hskip 7.5mm  between $\hat{\bold{x}}$ and $\bold{x}^*$.	
\vskip 2mm
\hrule}
\end{table}

To extend the algorithm in Table~\ref{algor1} from deterministic ODEs to 
stochastic models requires substantial modifications~\cite{Toni2009}. 
One main obstacle is the step (a2) which requires repeatedly generating 
the likelihood function $\mathcal{L}(\bold{k}^* | \hat{\bold{x}})$, as 
the outcome of stochastic models. In this case, a modeller 
must either apply statistical analysis to approximate the
likelihood~\cite{Reinker2006}, or use the Gillespie SSA to estimate
it~\cite{Tian2007}. Consequently, the algorithms are computationally 
intensive and do not scale well to problems of realistic size 
and complexity. To avoid this problem, the TPA uses  the tensor formalism
to separate the simulation part from the parameter inference. The parameter 
estimation is performed on the tensor data obtained by methods described
above (see Table~\ref{tabsimdetails}). The algorithm used for 
the TPA parameter estimation is given in Table~\ref{algor2}. 
The distance function $d(\hat{\bold{x}}, \bold{x}^*)$ is replaced 
with a distance between summary statistics, $\hat{S}$ and $S^*$, 
which describe the average behaviour and the characteristics of the 
system noise. The steps (b1) and (b3) are similar to steps (a1) 
and (a3) under the ODE settings, and a variety of existing methods 
can be extended directly to stochastic settings. The newly introduced 
step (b0) is executed only once during the parameter estimation.
Steps (b1)--(b3) are then repeated until convergence. 
Step (b2) only requires manipulation of tensor data, of which the 
computational overhead is comparable to solving an ODE.

\smallskip

\begin{table}[t]
\caption{{\modify{\it An algorithm for the tensor-structured 
parameter estimation.}}}
\label{algor2}
{\modify
\vskip 2mm
\hrule
\vskip 2mm
{\bf (b0)}  Compute the stationary distribution $p(\bold{x} | \bold{k})$ 
for all considered combinations of $\bold{x} \in \Omega_\bold{x}$ and 
$\bold{k} \in \Omega_\bold{k}$; and \\
\rule{0pt}{1pt} \hskip 7.5mm  
store $p(\bold{x} | \bold{k})$ in tensor data.\\
{\bf (b1)} Generate a candidate parameter vector 
$\bold{k}^* \in \Omega_\bold{k}$.\\
{\bf (b2)} Extract the stationary distribution 
$p(\bold{x} | \bold{k}^*)$ from the tensor-structured 
data $p(\bold{x} | \bold{k})$, and \\
\rule{0pt}{1pt} \hskip 7.5mm compute the summary statistics 
$S^* \equiv S^*(p)$.\\
{\bf (b3)}  Compare the model prediction $S^*$ with the statistics 
$\hat{S}$ obtained from experimental data, using \\ 
\rule{0pt}{1pt} \hskip 7.5mm distance function $J(\hat{S}, S^*)$ 
and tolerance $\varepsilon$.
If $J(\hat{S}, S^*)  < \varepsilon$, then accept $\bold{k}^*$. \\
\rule{0pt}{1pt} \hskip 7.5mm 
The tolerance $\varepsilon > 0$ is the desired level of agreement between 
$\hat{S}$ and $S^*$.
\vskip 2mm
\hrule}
\end{table}

\begin{figure}[b!]%
%\centerline{
\noindent
\rule{0pt}{0pt} \hskip 1mm
\raise 57mm \hbox{(a)}
\hskip -4mm
\includegraphics[width=.45\textwidth]{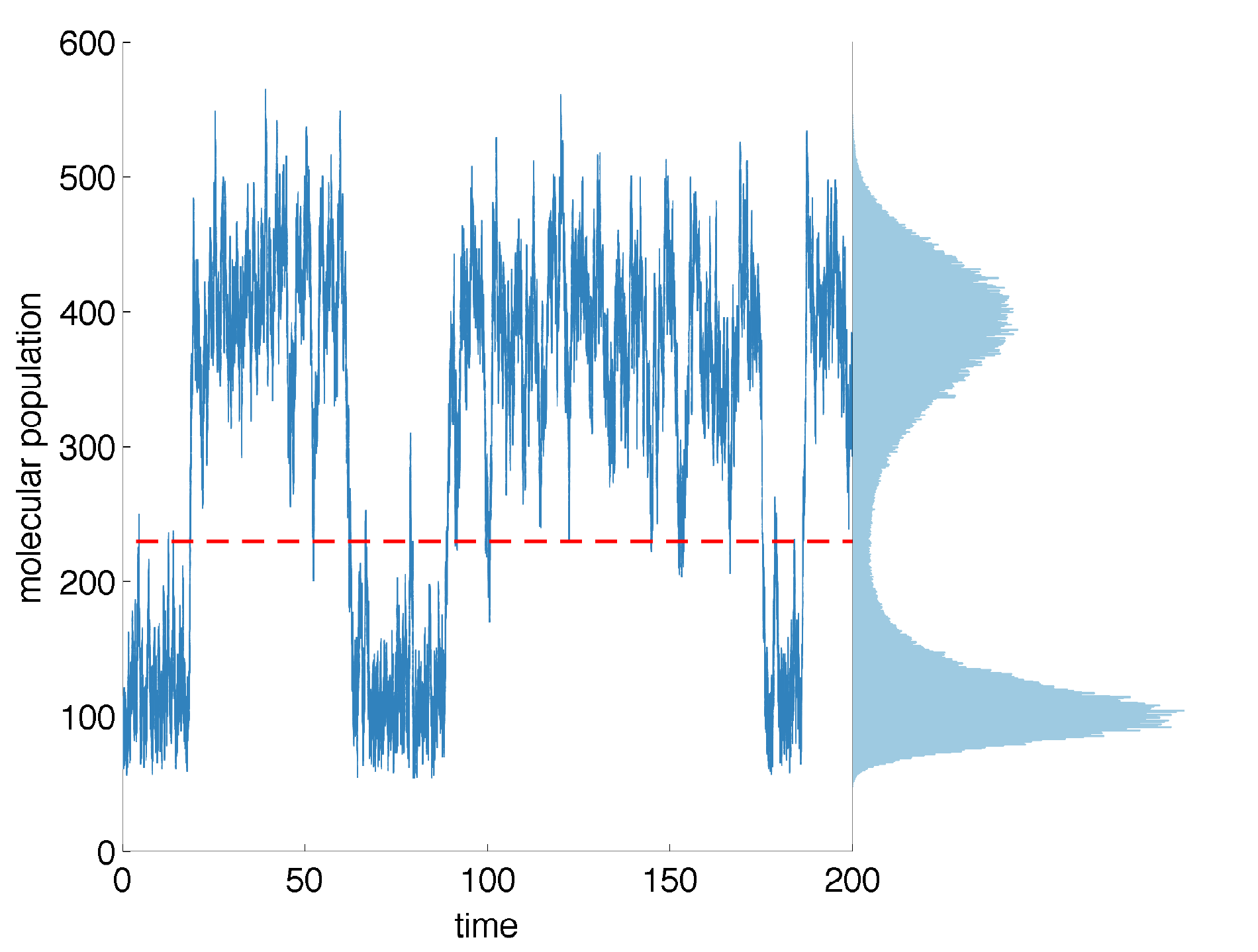}
\rule{0pt}{0pt} \hskip 1mm
\raise 57mm \hbox{(b)}
\hskip -7mm
\raise 4mm \hbox{\includegraphics[scale=.55]{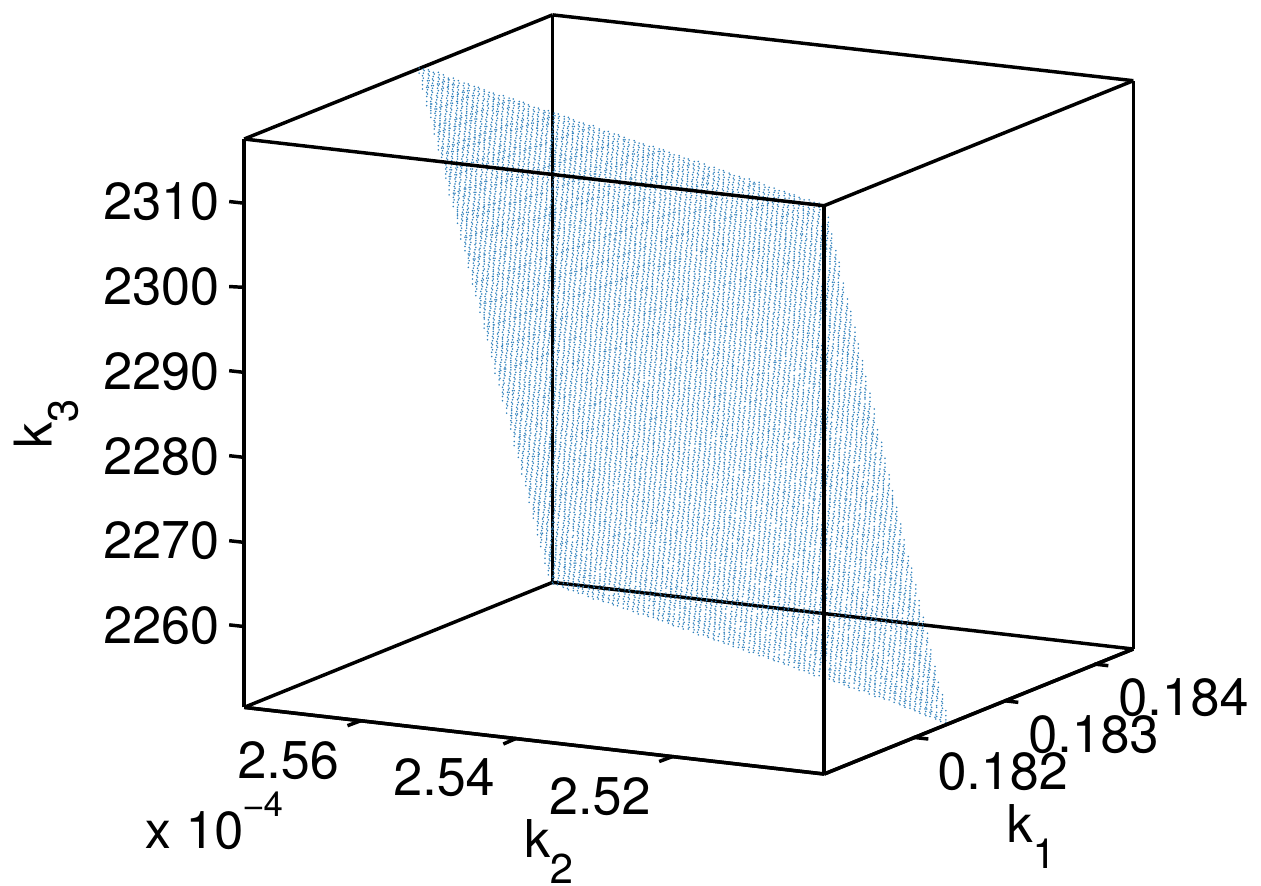}}
\\
\caption{%
(a) {\it A short segment of the time series data 
and the histogram for the Schl\"{o}gl reaction 
system generated by a long-time stochastic simulation.
The dashed line corresponds to the threshold $230$ which
is used to separate the two macroscopic states of 
this bistable system.}
(b) {\it The triplets of parameters $[k_1,k_2,k_3]$
for which the splitting probability $(\ref{splitprob})$
is equal to $\hat{S}=47.61\% \pm 5\%$. 
The value of $k_4$ is fixed at its true value.}
\label{fig1}
}
\end{figure}

\subsection{An example of parameter estimation}

We consider that the distance measure $J(\hat{S}, S^*)$ 
in Table~\ref{algor2} is defined using a moment matching 
procedure~\cite{Lillacci2010,Zechner2012}:
\begin{equation}
J(\hat{S}, S^*)
= 
\sum_{i_1, \ldots, i_N=1}^{L} \beta_{i_1, \ldots, i_N}
\left( 
\frac{\hat{\mu}_{[i_1, \ldots, i_N]} - \mu_{[i_1,\ldots, i_N]}(\mathbf{k}^*)}
{\hat{\mu}_{[i_1,\ldots, i_N]}} \right)^2,
\label{momdist}
\end{equation}
where $\hat{\mu}_{[i_1,\ldots, i_N]}$ is the $(i_1, \ldots, i_N)$-th order 
empirical raw moment, $\mu_{[i_1,\ldots, i_N]}(\mathbf{k}^*)$ is the 
corresponding moment derived from $p(\mathbf{x}|\mathbf{k}^*)$ and 
$L$ denotes the upper bound for the moment order. 
The weights, $\beta_{i_1, \ldots, i_N}$, 
can be chosen by modellers to attribute different relative importances 
to moments. Empirical moments are 
estimated from samples $\hat{x}_{d,\ell}$, $d = 1,2,\ldots, N$, 
$\ell = 1, 2, \ldots, n_{\hat{\mu}}$, by 
\begin{equation}
\hat{\mu}_{[i_1,\ldots, i_N]} 
= \frac{1}{n_{\hat{\mu}}} 
\sum _{\ell = 1}^{n_{\hat{\mu}}} 
\hat{x}_{1,\ell}^{i_1} \cdots \hat{x}_{N,\ell}^{i_N},
\label{empmom}
\end{equation}
where $n_{\hat{\mu}}$ is the number of samples.
Moments of the model output are computed as 
\begin{equation}
\mu_{[i_1,\ldots, i_N]}(\mathbf{k}^*) 
= \int_{\Omega_{\mathbf{x}}} x_1^{i_1} \cdots x_N^{i_N} \, 
p\left(\mathbf{x} | \mathbf{k}^*\right) \mathrm{d} \mathbf{x}. 
\label{momentsofpxk}
\end{equation}
We show, in \emph{SI Appendix} S1.4, that it is possible to directly compute 
different orders of moments, $\mu_{[i_1,\ldots, i_N]}(\mathbf{k}^*)$, using 
the representation (\ref{tensorrep}) with $O(N)$ complexity.

We illustrate the tensor-structured parameter estimation using the Schl\"{o}gl 
chemical system~\cite{schlogl1972chemical}, which is written 
for $N=1$ molecular 
species and has $M=4$ reaction rate constants $k_i$, $i=1,2,3,4$. 
A detailed description of this system is provided in \emph{SI Appendix} S2.1.
We prescribe true parameter values as $k_1 = 2.5 \times 10^{-4}$, 
$k_2 = 0.18$, $k_3 = 2250$ and $k_4 = 37.5$, and use a long-time 
stochastic simulation to generate a time series as pseudo-experimental data 
(for a short segment, see Figure~\ref{fig1}(a)). 
These pseudo-experimental data are then used 
for estimating the first three empirical moments $\hat{\mu}_i$, 
$i = 1, 2, 3$, using (\ref{empmom}). 
While the moments of the model output, $\mu_{i}(\mathbf{k})$, 
$i = 1, 2, 3$, are derived from the tensor-structured data 
$p(\mathbf{x}|\mathbf{k})$, computed using (\ref{tensorrep}). 
Moment matching is sensitive to the choice of weights~\cite{Zechner2012}. 
However, for the sake of simplicity we choose 
the weights $\beta_i$, $i = 1,2, 3$, 
in a way that the contributions of the different orders of moments are 
of similar magnitude within the parameter space.
Having the stationary distribution stored in the tensor format
(\ref{tensorrep}), we can then efficiently iterate steps (b1)--(b3) 
in Table~\ref{algor2} to search for parameter 
values that produce adequate fit to the samples using the measure
given in equation (\ref{momdist}). 
We consider $\varepsilon = 0.25\%$ and visualise in 
Figure~\ref{fig2} the admissible parameter values satisfying 
$J(\hat{S}, S^*) < \varepsilon$.

\begin{figure}[t]%
\centerline{
%\noindent
\rule{0pt}{0pt} \hskip 1mm
\raise 3.7cm \hbox{(a)}
\hskip -8mm
\includegraphics[scale=.45]{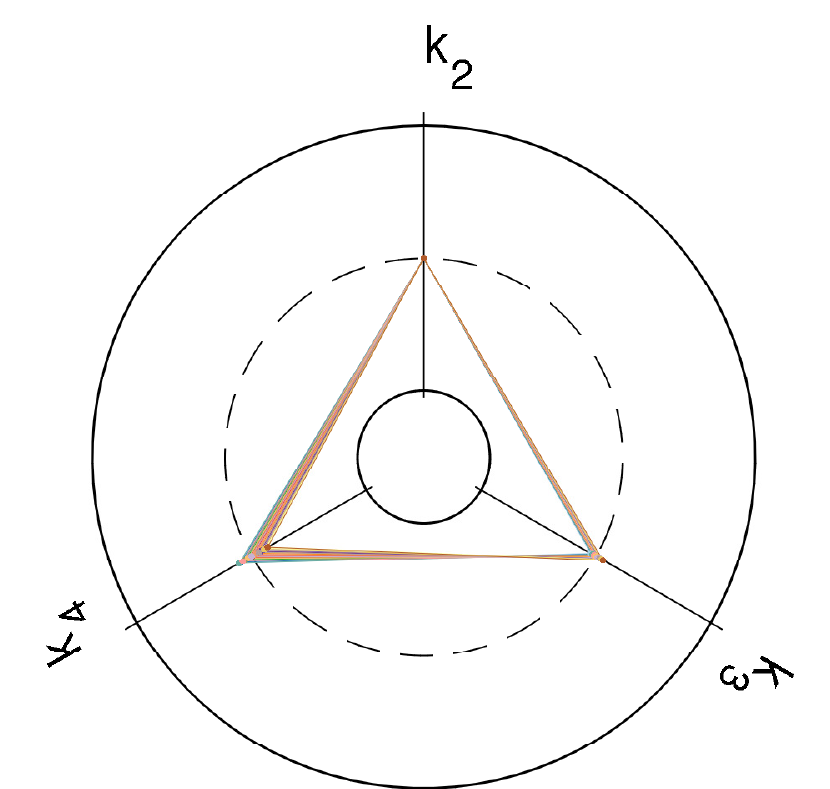}
\hskip 6mm
\raise 3.7cm \hbox{(b)}
\hskip -11mm
\includegraphics[scale=.45]{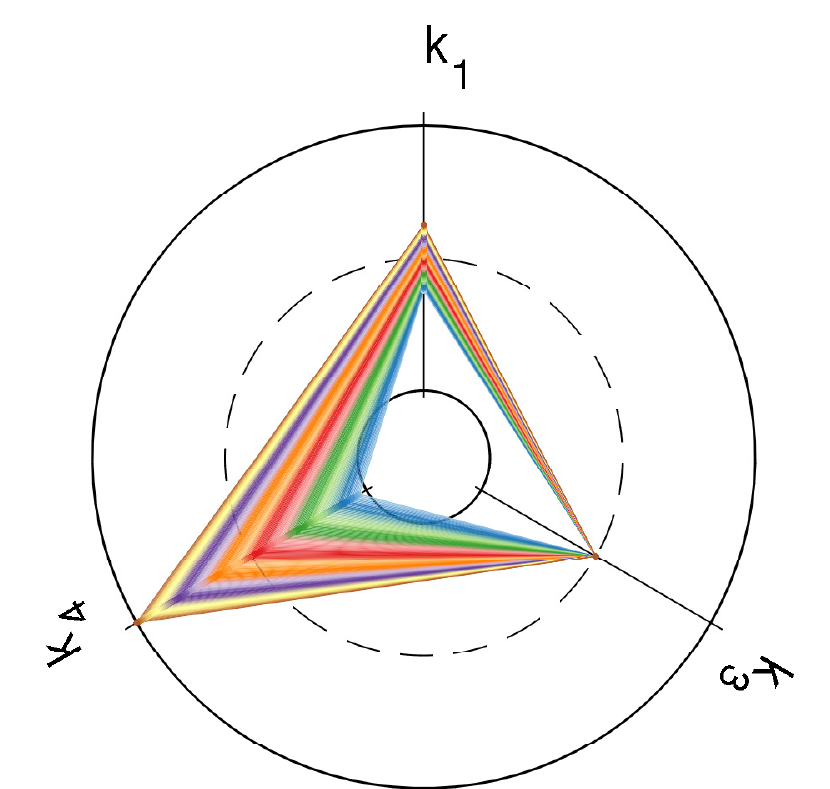}
\hskip 6mm
\raise 3.7cm \hbox{(c)}
\hskip -11mm
\includegraphics[scale=.45]{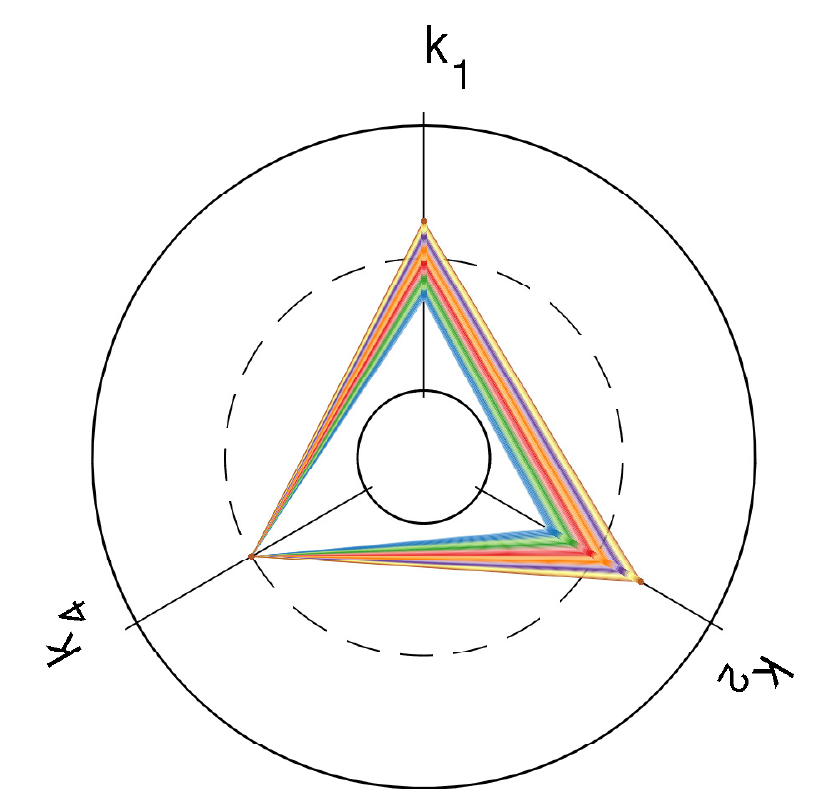}
\hskip 6mm
\raise 3.7cm \hbox{(d)}
\hskip -11mm
\includegraphics[scale=.45]{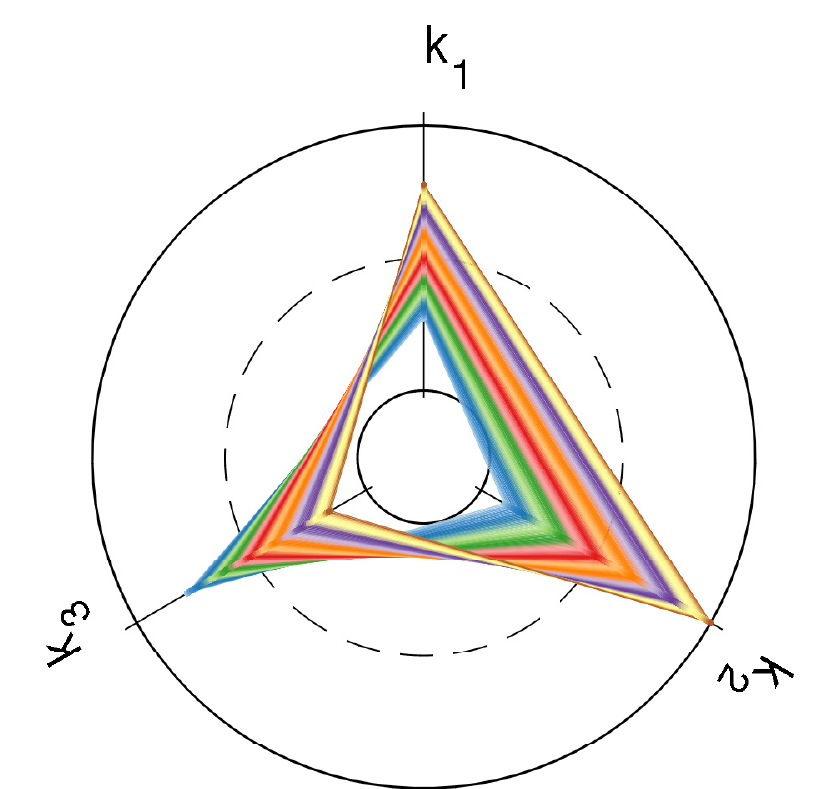}
}
\caption{{\it Circular representation}~\cite{von2000segment} 
{\it of estimated 
parameter combinations for the Schl\"ogl model. Each spoke 
represents the corresponding 
parameter range listed in Table}~S4. {\it The true parameter values are specified by the intersection points between the spokes and the dashed 
circle. Each triangle (or polygon in general) of a fixed colour corresponds 
to one admissible parameter set with $\varepsilon = 0.25\%$. 
Each panel} (a)--(d) {\it shows the situation with one parameter 
fixed at its true value, namely:}
(a) $k_1$ {\it is fixed};
(b) $k_2$ {\it is fixed};
(c) $k_3$ {\it is fixed};
(d) $k_4$ {\it is fixed}.
}
\label{fig2}
\end{figure}

The summary statistics $\hat{S}$ are not restricted to 
lower order moments. The TPA can efficiently evaluate different 
choices of the summary statistics, because of the simplicity and 
generality of separable representation (\ref{tensorrep}). 
For example, if one can experimentally measure the probability 
that the system stays in each of the two states of the bistable 
system, then distance measure $J(\hat{S}, S^*)$ can be based on the 
probability of finding the system within a particular part of the state 
space $\Omega_\mathbf{x}$. We show in \emph{SI Appendix} S1.4 
that such quantity can also be estimated in the tensor format efficiently 
with $O(N)$ complexity. Considering the Schl\"{o}gl model, we estimate the 
probability that the system stays in the state with less molecules by
\begin{equation}
S
=
{\mathbb P}(x \leq 230), 
\label{splitprob}
\end{equation}
where ${\mathbb P}$ denotes the probability and 
the threshold $230$ separates the two macroscopic states 
of the Schl\"ogl system, see the dashed line in Figure~\ref{fig1}(a).
The splitting probability (\ref{splitprob}) can be estimated
using long time simulation of the Schl\"ogl system as the fraction 
of states which are less or equal than 230 and is equal
to $\hat{S} = 47.61\%$ for our true parameter values.
Figure~\ref{fig1}(b) shows the set of admissible parameters within 
the parameter space $\Omega_\mathbf{k}$ whose values provide 
desired agreement on the splitting probability (\ref{splitprob})
with tolerance $\varepsilon = 5\%$, i.e. we use
$$
J(\hat{S}, S^*)
=
| \hat{S} - S^* |
$$
in the algorithm given in Table~\ref{algor2}, where
$S^*$ is computed using (\ref{tensorrep}) and (\ref{splitprob}).

\subsection{Identifiability}

One challenge of mathematical modelling of GRNs
is whether unique parameter values can be determined from available data. 
This is known as the problem of identifiability.  Inappropriate choice of the 
distance measure may yield ranges of parameter values with equally good 
fit, i.e. the parameters being not 
identifiable~\cite{Gutenkunst2007}. 
Here, we illustrate the tensor-structured identifiability analysis 
of the deterministic and stochastic models of the Schl\"{o}gl
chemical system. We plot the distance function against two parameter 
pairs, rate constants $k_1$-$k_3$ and $k_2$-$k_4$, in
Figure~\ref{fig3}. From the colour map, we see that 
the distance function (\ref{momdist}) possesses
a well distinguishable global minimum at the true values
($k_1 = 2.5 \times 10^{-4}$, $k_2 = 0.18$, $k_3 = 2250$
and $k_4 = 37.5$). This indicates that
the stochastic model is identifiable in both cases. In the 
deterministic scenario, the Schl\"ogl system 
loses its identifiability. When the distance function 
(\ref{momdist}) 
only fits the mean concentration, the minimal values are attained 
on a curve in the 2D parameter 
space (the distance function is indicated by blue contour 
lines in Figure~\ref{fig3}). 
Stochastic models are advantageous in model identifiability, 
because they can be parametrized using a wider class of statistical 
properties (typically, $K$ quantities are needed to 
estimate $K$ reaction rate constants for mass-action reaction
systems). The TPA enables efficient 
and direct evaluation of $J(\hat{S}, S^*)$
all over the parameter space in a single computation
by using the representation (\ref{tensorrep}).

Figure~\ref{fig3} also reveals the differences between the model 
responses to parameter perturbations.
The green contour lines show the landscape of $J(\hat{S}, S^*)$ for 
the stochastic model using only 
the mean values, i.e., $L =1$ in (\ref{momdist}). 
The minimum is attained 
on a straight line, representing another non-identifiable situation. 
This line (green) has a different direction than the line obtained
for the deterministic model (blue). In particular,  
this example illustrates that the parameter values 
estimated from deterministic models
do not give good approximation of both average behaviour and 
the noise level when they are used
in stochastic models~\cite{Wilkinson2009}.
}

\begin{figure}[t]
\centerline{
\noindent
\hskip 1mm
\raise 6.2cm \hbox{(a)}
\hskip -7mm
\includegraphics[width=.46\textwidth]{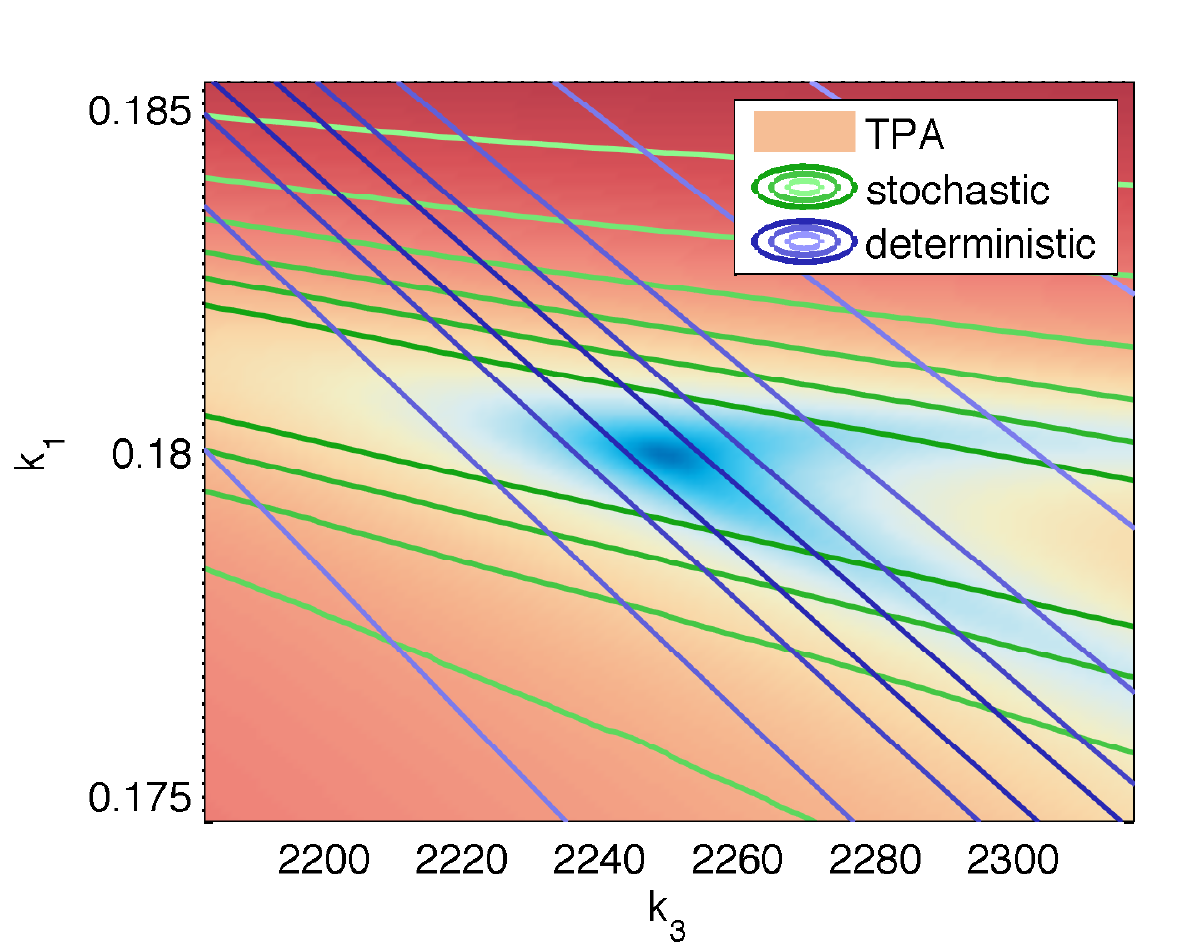}
\hskip 8mm
\raise 6.2cm \hbox{(b)}
\hskip -8mm
\includegraphics[width=.46\textwidth]{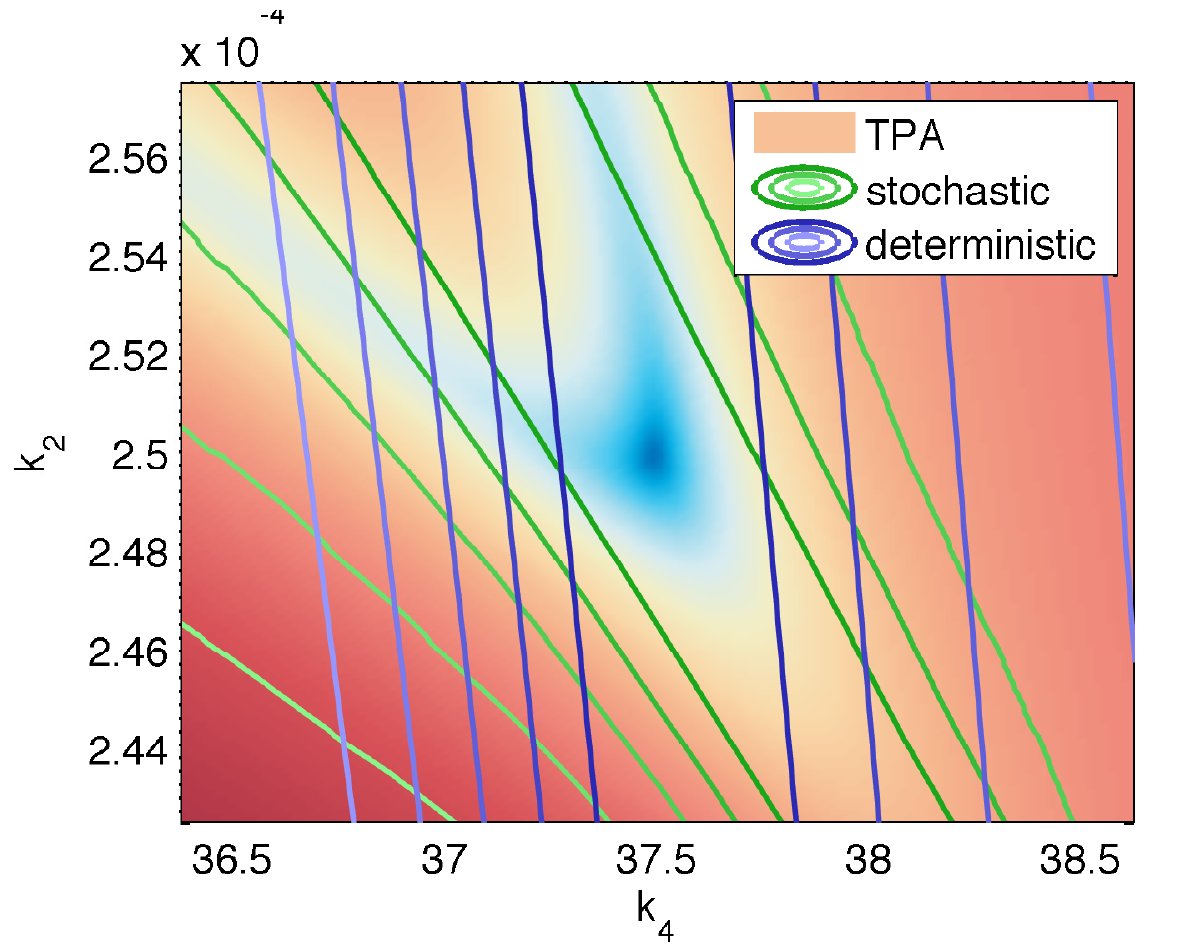}
}
\caption{{\it Parameter identifiability analysis of the Schl\"{o}gl reaction 
system. } \hfill\break
(a)~{\it Estimation and identifiability of parameters $k_1$ and $k_3$ 
with $k_2$ and $k_4$ fixed at their true values. 
The colour scale corresponds to values of the distance function }
(\ref{momdist}) {\it with $L=3$, i.e., the first three moments are
compared. The green and blue contour lines indicate the distance 
functional } (\ref{momdist}) {\it with the first moment 
(mean value) only. Green corresponds to the stochastic model and blue 
to the deterministic model.} \hfill\break
(b)~{\it Estimation and identifiability of parameters $k_2$ and $k_4$ 
with $k_1$ and $k_3$ fixed at their true values. The same quantities
as in panel} (a) {\it are plotted.}
\label{fig3}
}
\end{figure}

\section{Bifurcation analysis}
\label{secbifan}

\begin{figure*}[t]
\noindent
\raise 4.65cm 
\hbox{(a)}
\hskip -5mm
\includegraphics[width=.2\textwidth]{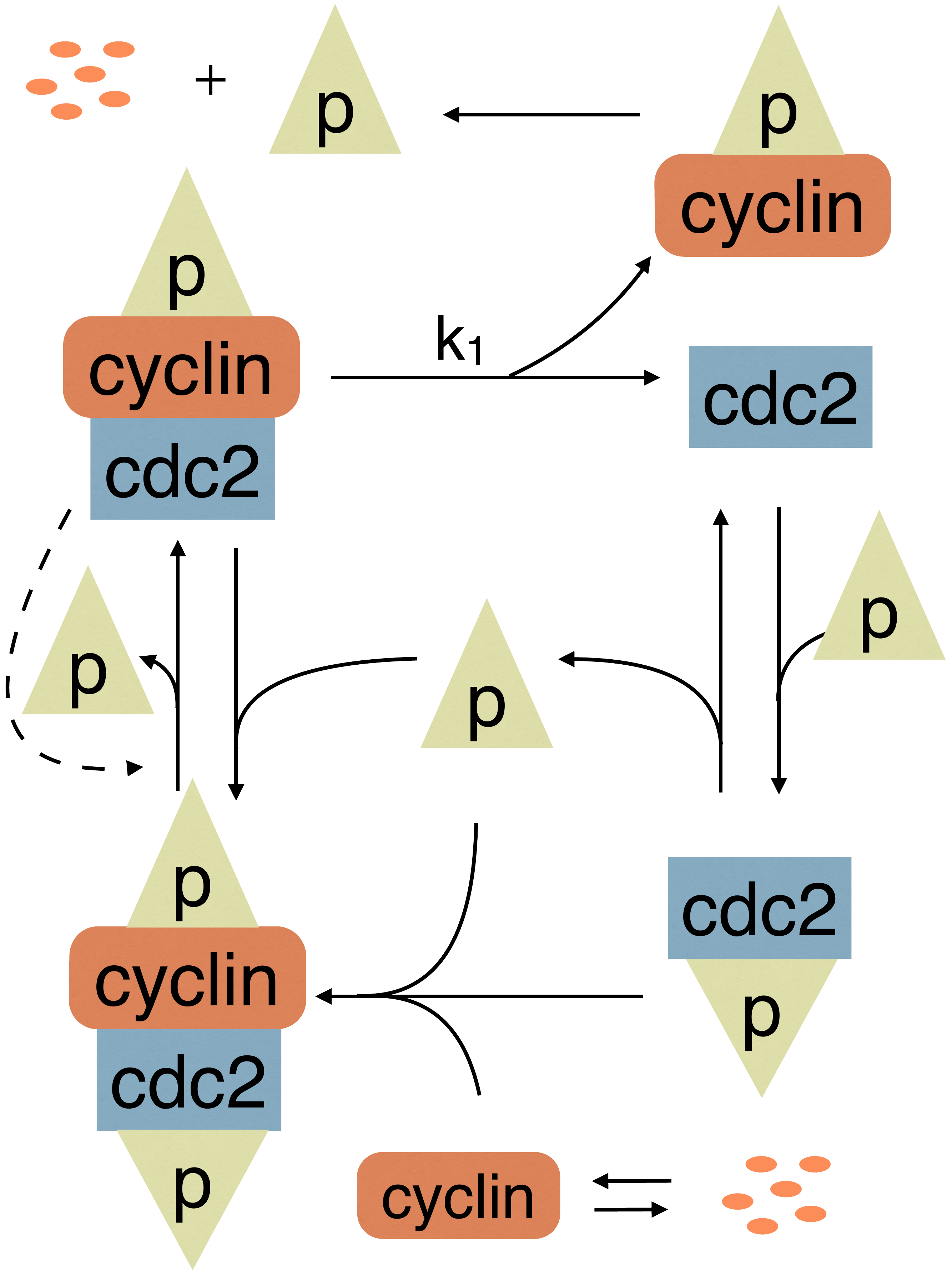}
\hskip 10mm
\raise 4.65cm
\hbox{(b)}
\hskip -5mm
\includegraphics[width=.33\textwidth]{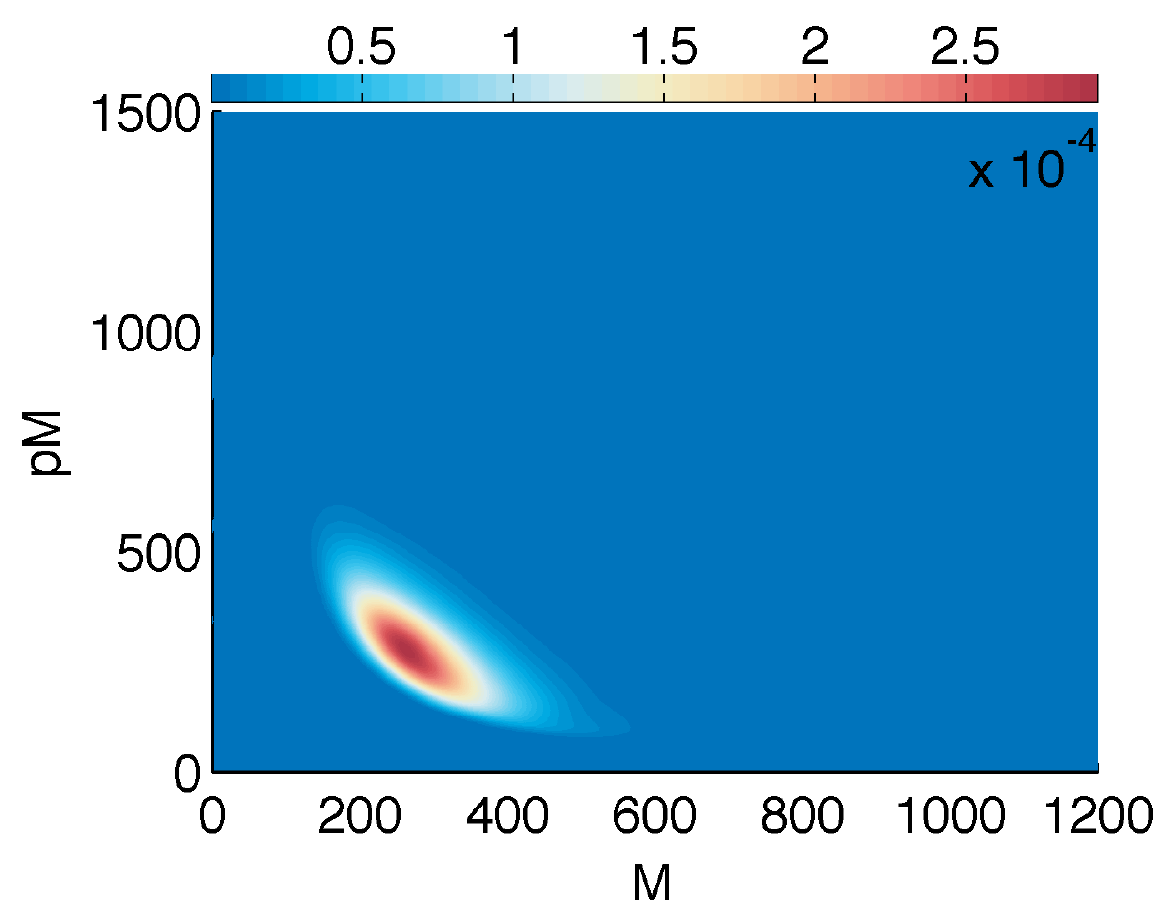}
\hskip 8mm
\raise 4.65cm 
\hbox{(c)}
\hskip -8mm
\includegraphics[width=.33\textwidth]{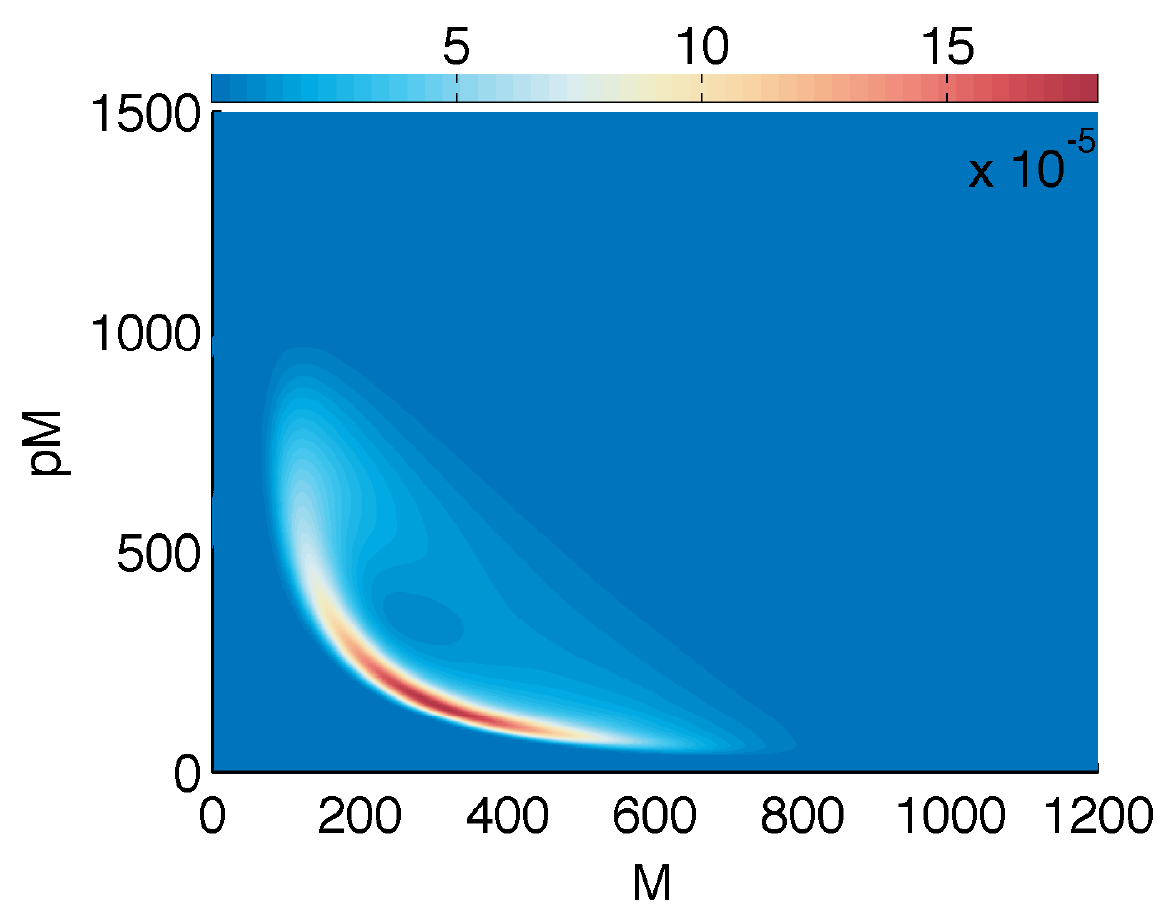}
\caption{{\it 
Bifurcation analysis of the stochastic cell cycle model.} \hfill\break
(a)~{\it Schematic description of the cyclin-cdc$2$ interactions. 
Free cyclin molecules combine rapidly with phosphorylated cdc2, to 
form the dimer MPF (cdc2-cyclin-p), which is immediately inactivated 
by phosphorylation process. The inactive MPF (p-cdc2-cyclin-p) can 
be converted to active MPF by autocatalytic dephosphorylation. The 
active MPF in excess breaks down into cdc2 molecules and phosphorylated 
cyclin, which is later subject to proteolysis. Finally, cdc2 
is phosphorylated to repeat the cycle.} \hfill\break
(b)~{\it Joint stationary distribution of cdc$2$-cyclin-p (M) and 
p-cdc$2$-cyclin-p (pM) 
plotted at the deterministic bifurcation point ($k_1 =0.2694$).} 
\hfill\break
(c)~{\it Joint stationary distribution of cdc$2$-cyclin-p (M) and 
p-cdc$2$-cyclin-p (pM) 
plotted for $k_1 = 0.3032$.}}
\label{fig4}
\end{figure*}

Bifurcation is defined as a qualitative transformation in the behaviour 
of the system as a result of the continuous change in model parameters.
Bifurcation analysis of ODE systems has been used to understand the 
properties of deterministic models of biological systems, including 
models of cell cycle~\cite{Tyson:2002:DCC} and circadian 
rhythms~\cite{Scheper:1999:MMI}. Software packages, implementing 
numerical bifurcation methods for ODE systems, have also been presented
in the literature \cite{Doedel:1997:ACB,Dhooge:2003:MMP}, but computational
methods for bifurcation analysis of corresponding stochastic models
are still in development \cite{Erban:2009:ASC}.
{\modify
Here, we use the tensor-structured data $p(\mathbf{x}|\mathbf{k})$ 
given by (\ref{tensorrep}) for a model of fission yeast cell cycle 
control developed by Tyson~\cite{tyson1991modeling}, and perform the
tensor-structured bifurcation analysis on the tensor data.}
The interaction of cyclin-cdc2 in the Tyson model is illustrated in 
Figure~\ref{fig4}(a). Reactions and parameter values 
are given in \emph{SI Appendix} S2.2.

The parameter $k_1$, indicating the breakdown of the active 
M-phase-promoting factor (MPF), is chosen as the bifurcation parameter.
The analysis of the corresponding ODE model reveals that 
the system displays a stable steady state when $k_1$ is at its 
low values, which describes the metaphase arrest of unfertilized 
eggs~\cite{hunt1989under}. On the other hand, the ODE model
is driven into rapid cell cycling exhibiting 
oscillations when $k_1$ increases~\cite{tyson1991modeling}.
The ODE cell cycle model has a bifurcation point at $k_1 =0.2694$, where 
a limit cycle appears \cite{tyson1991modeling}. In our TPA computations,
we study the behaviour of the stochastic model for the values of
$k_1$ which are close to the deterministic bifurcation point.
We observe that the steady state distribution changes
from a unimodal shape (Figure~\ref{fig4}(b))
to a probability distribution with a ``doughnut-shaped" region of high 
probability (Figure~\ref{fig4}(c)) at $k_1 = 0.3032$.
In particular, the stochastic bifurcation appears for higher values of $k_1$ 
than the deterministic bifurcation. 

\begin{figure*}[t]
\noindent
\rule{0pt}{0pt} \hskip 1mm
\raise 3.1cm \hbox{(a)}
\hskip -10mm
\includegraphics[scale=.4]{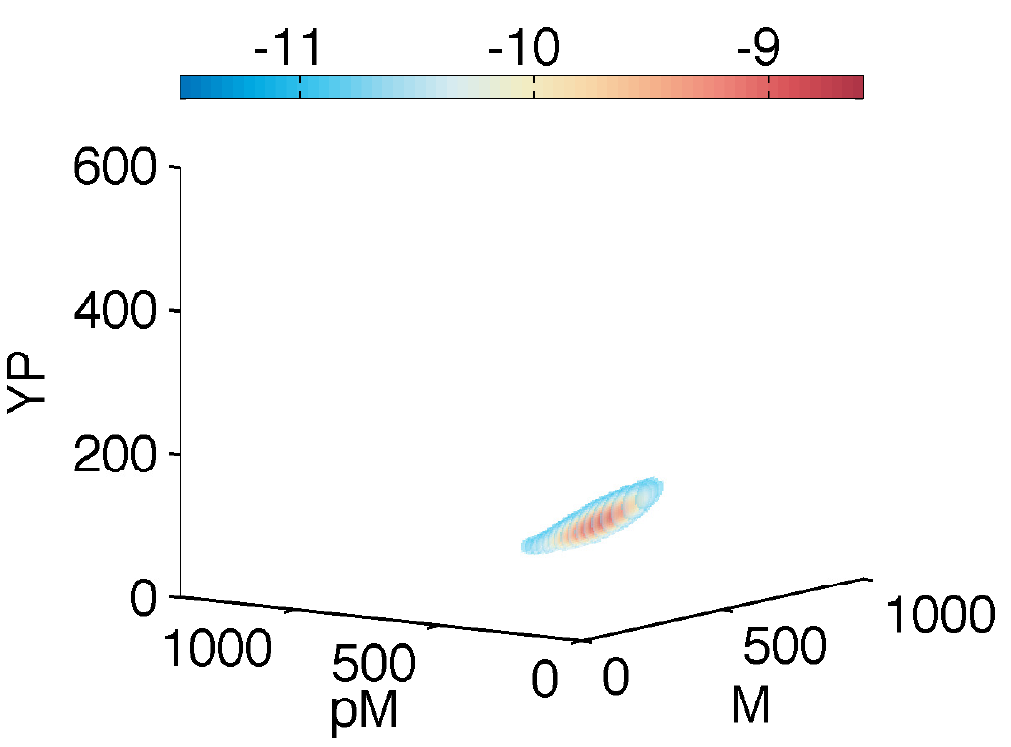}
\hskip 5mm
\raise 3.1cm \hbox{(b)}
\hskip -10mm
\includegraphics[scale=.4]{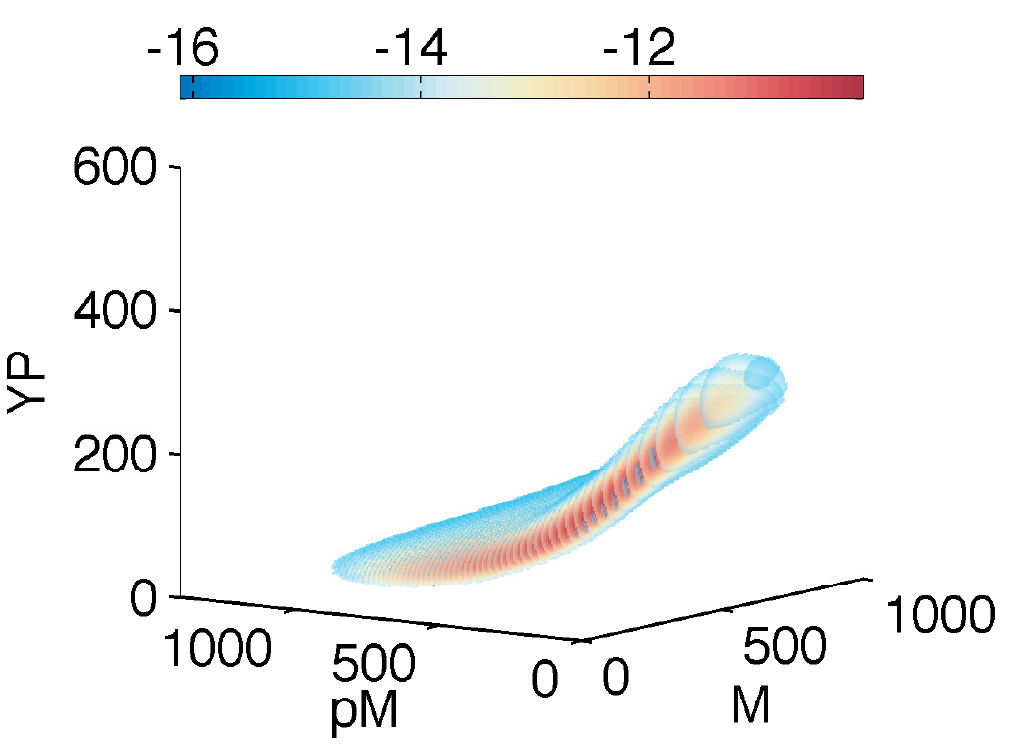}
\hskip 5mm
\raise 3.1cm \hbox{(c)}
\hskip -10mm
\includegraphics[scale=.4]{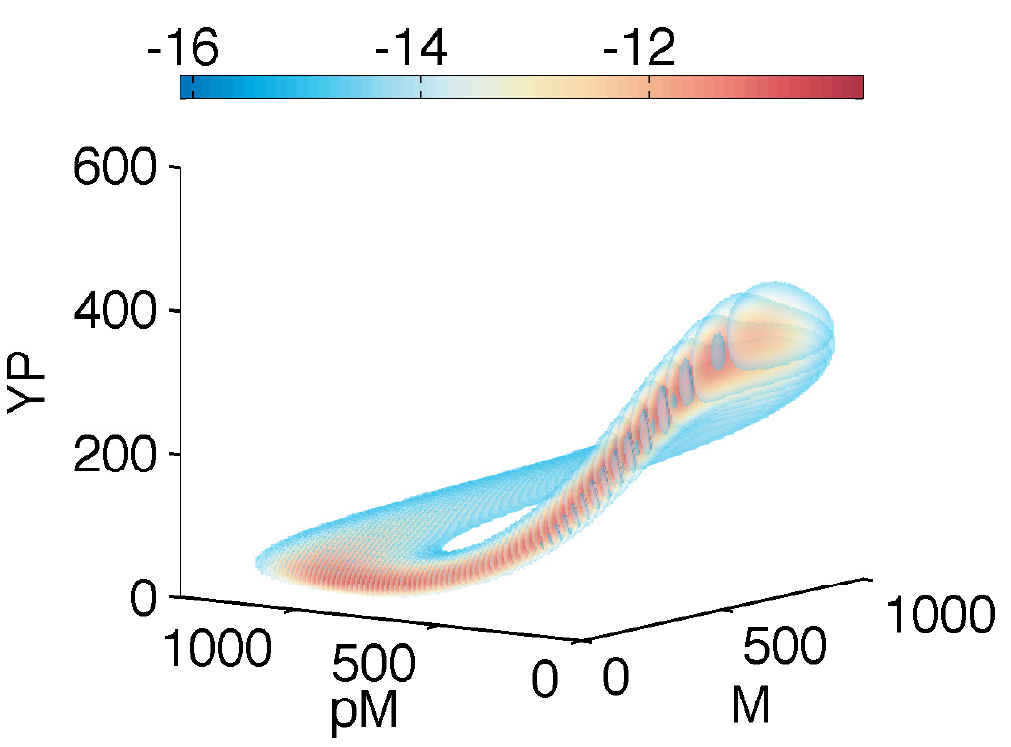}
\hskip 5mm
\raise 3.1cm \hbox{(d)}
\hskip -10mm
\includegraphics[scale=.4]{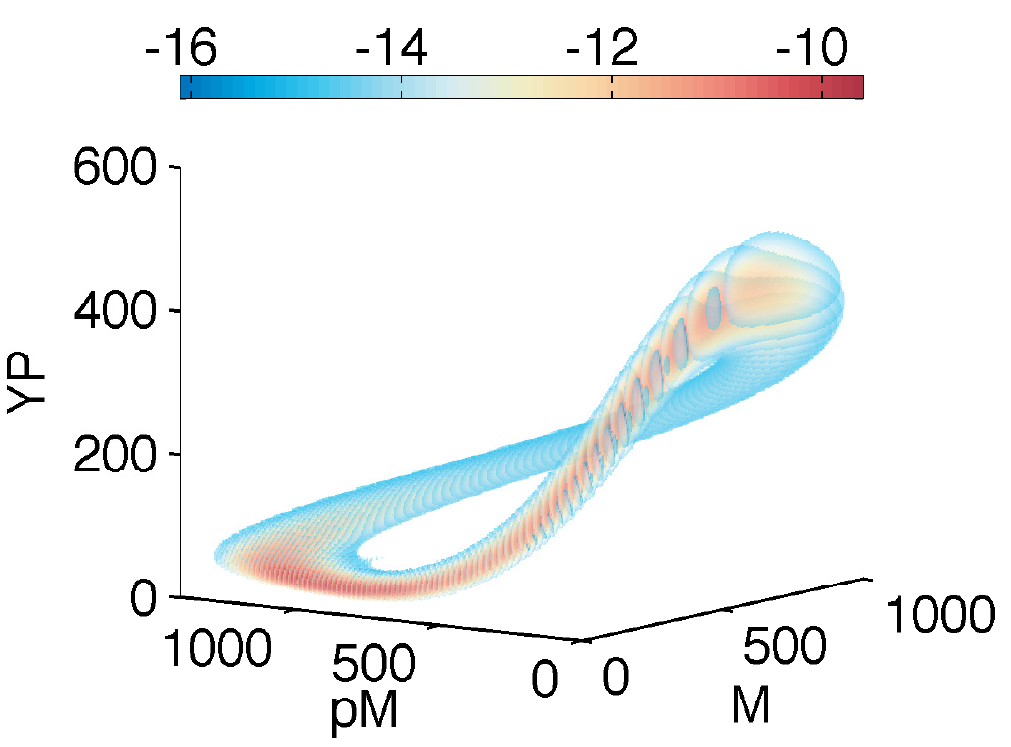} \\
\vskip 2mm
\noindent
\rule{0pt}{0pt} \hskip 1mm
\raise 3.1cm \hbox{(e)}
\hskip -10mm
\includegraphics[scale=.4]{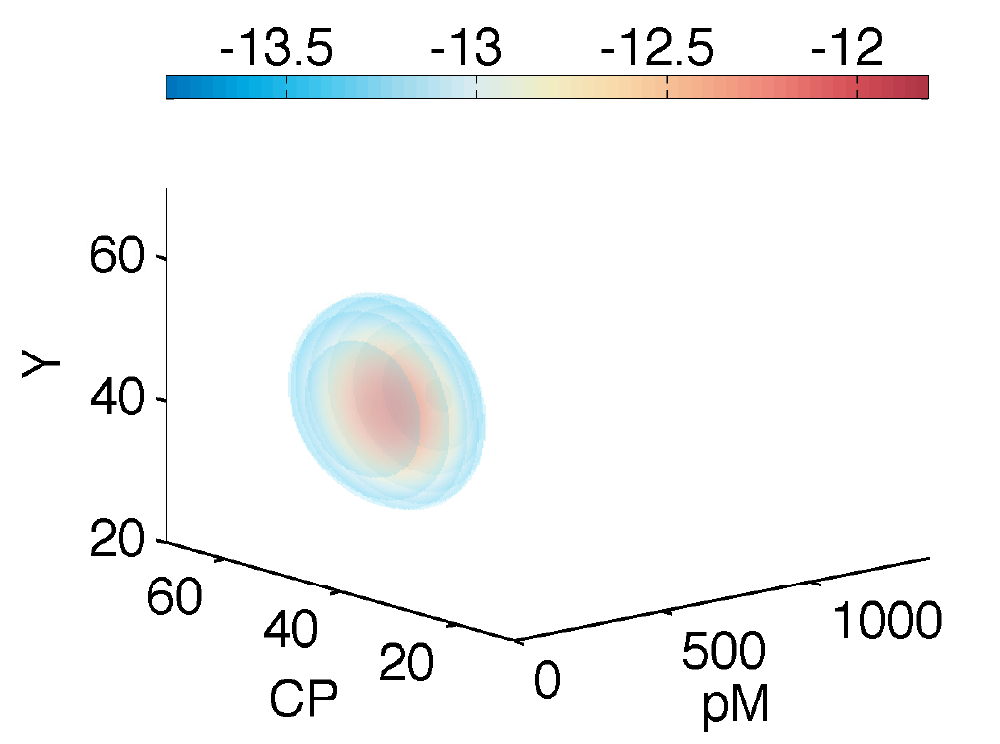}
\hskip 5mm
\raise 3.1cm \hbox{(f)}
\hskip -10mm
\includegraphics[scale=.4]{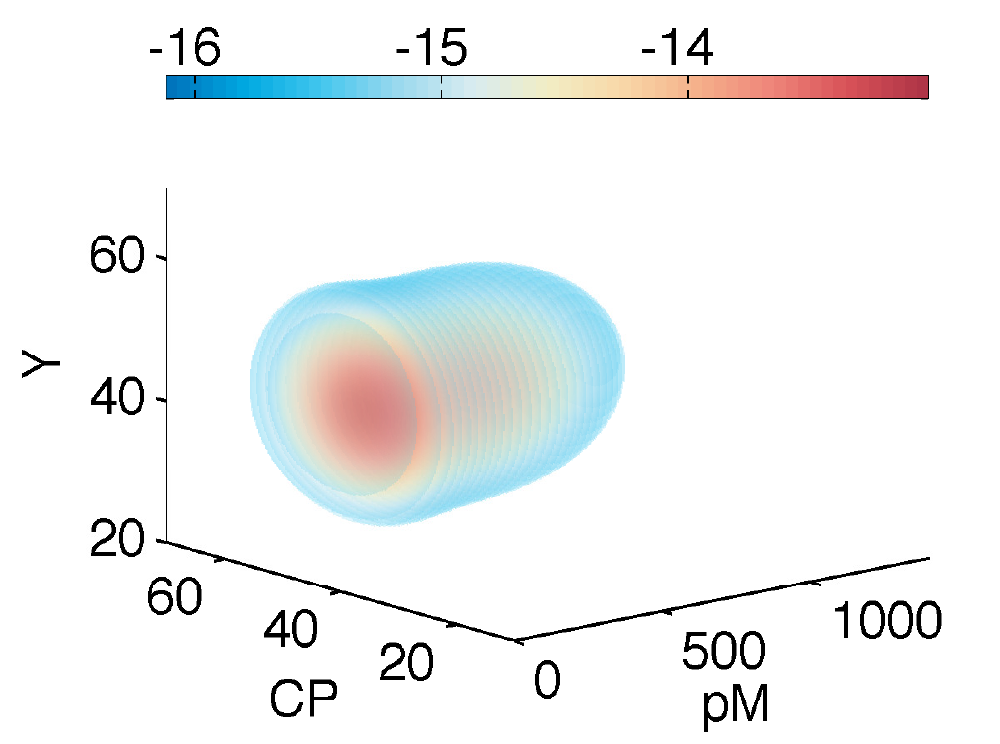}
\hskip 5mm
\raise 3.1cm \hbox{(g)}
\hskip -10mm
\includegraphics[scale=.4]{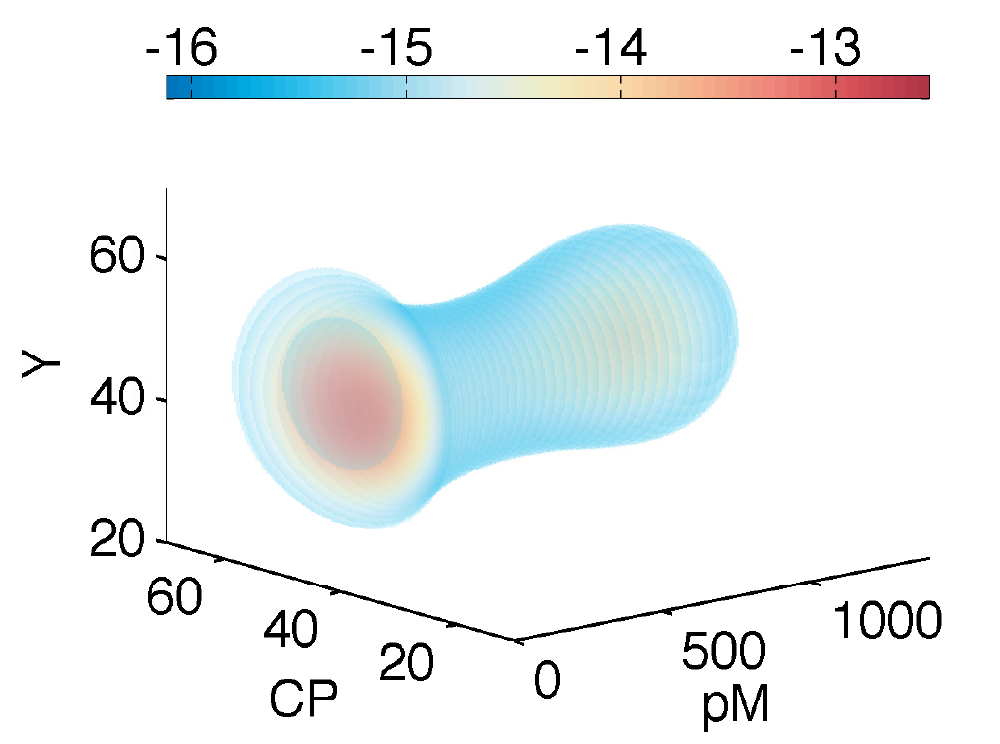}
\hskip 5mm
\raise 3.1cm \hbox{(h)}
\hskip -10mm
\includegraphics[scale=.4]{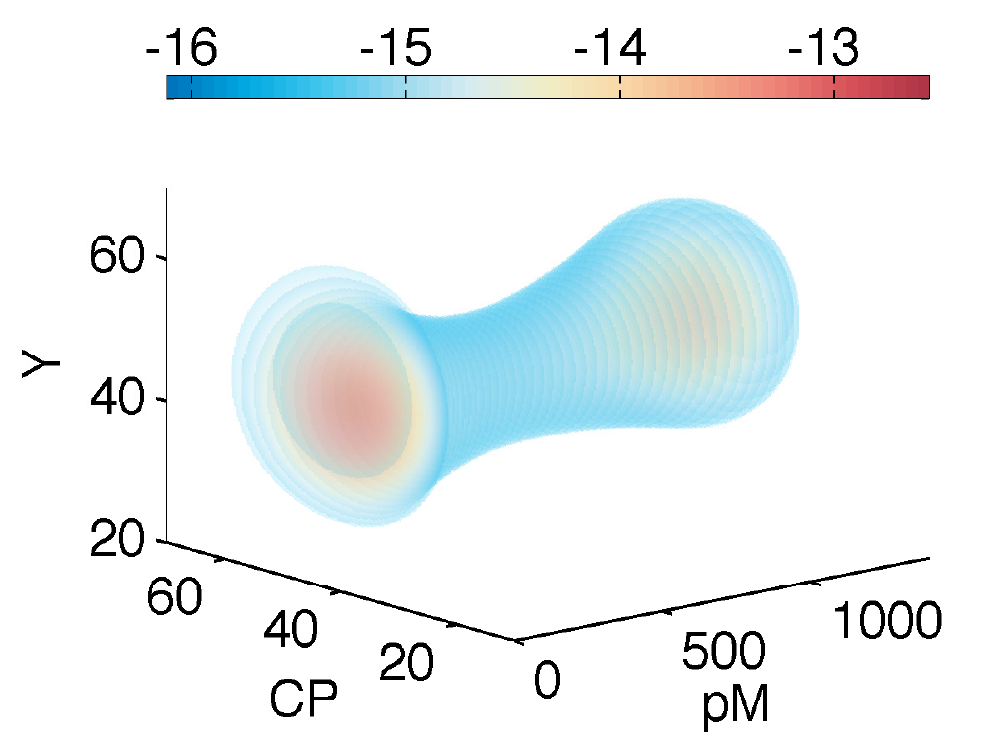}
\caption{
{\it Visualisation of the bifurcation structure of the stochastic cell cycle 
model.} \hfill\break
(a)--(d) {\it  Marginal steady state distributions of the phosphorylated 
cyclin (YP), the inactive MPF (pM) and the active MPF (M).} \hfill\break
(e)--(h) {\it  Marginal stationary distributions of the cyclin (Y), 
the phosphrylated cdc2 (CP) and the inactive MPF (pM). \hfill\break
Each column corresponds to the same value of the bifurcation 
parameter $k_1$, which from the left to the right are $0.25$, 
$0.3$, $0.35$ and $0.4$, respectively. All figures show 
$\log_{2}$ of the marginal steady state distribution 
for better visualisation.}}
\label{fig5}
\end{figure*}

In Figure~\ref{fig5}, we use the computed tensor-structured parametric 
probability distribution to visualise 
the stochastic bifurcation structure of the cell cycle model. 
As the bifurcation parameter $k_1$ increases, the expected oscillation 
tube is formed and amplified in the marginalised  YP-pM-M state 
space (see panels (a)--(d) of Figure~\ref{fig5}). 
In panels (e)--(h) of Figure~\ref{fig5}, the marginal distribution in 
the Y-CP-pM subspace is plotted. 
We see that it changes from a unimodal (Figure~\ref{fig5}(e)) to a bimodal
distribution (Figure~\ref{fig5}(f)). 
Cell cycle models have been studied in the deterministic context either 
as oscillatory~\cite{tyson1991modeling} 
or bistable~\cite{pomerening2003building,xiong2003positive}
systems. In Figure~\ref{fig5}, we see that the presented
stochastic cell cycle model can appear to have both oscillations 
and bimodality, when diffferent subsets of observables are
considered.

\section{Robustness analysis}
\label{secrobust}

GRNs are subject to extrinsic noise which is manifested 
by fluctuations of parameter values~\cite{paulsson2004summing}. 
This extrinsic noise originates from interactions of the modelled 
system with other  stochastic processes in the cell or its 
surrounding environment. 
We can naturally include extrinsic fluctuations under the 
tensor-structured framework.
For a GRN as in (\ref{reactionsystem}), we consider 
the copy numbers $X_1,X_2,\ldots,X_N$ as intrinsic variables 
and reaction rates $k_1,k_2,\ldots,k_M$ as extrinsic variables. 
Total stochasticity is quantified by the stationary distribution 
of the intrinsic variables, $p(\mathbf{x})$. 
We assume that the invariant probability density of extrinsic 
variables, $q ( \mathbf{k} )$, does not depend on the values 
of intrinsic variables $\mathbf{x}$.
Then the law of total probability implies that the stationary 
probability distribution of intrinsic variables is given by
\begin{equation}
p(\mathbf{x}) = \int_{\Omegak} p ( \mathbf{x}|\mathbf{k} ) 
\, q ( \mathbf{k}) \, \mathrm{d} \mathbf{k},
\label{probdistint}
\end{equation}
where $\Omegak$ is the parameter space and 
$p ( \mathbf{x}|\mathbf{k} )$ 
represents the invariant density of intrinsic variables conditioned 
on constant values of kinetic parameters, see the definition 
below equation~(\ref{reactionsystem}).
If distributions $q ( \mathbf{k} )$ of extrinsic variables 
can be determined from high quality experimental data then the 
stationary density can be computed directly by 
(\ref{probdistint}). If not, the TPA framework
enables to test the behaviour of GRNs 
for different hypothesis about the distribution of the extrinsic 
variables. The advantage of the TPA is that it efficiently computes 
the high-dimensional integrals in (\ref{probdistint}),  
see \emph{SI Appendix} S1.4.

\begin{figure}[h]
\noindent
\raise 5.5cm
\hbox{(a)}
\hskip -1.4mm
\includegraphics[width=.25\textwidth]{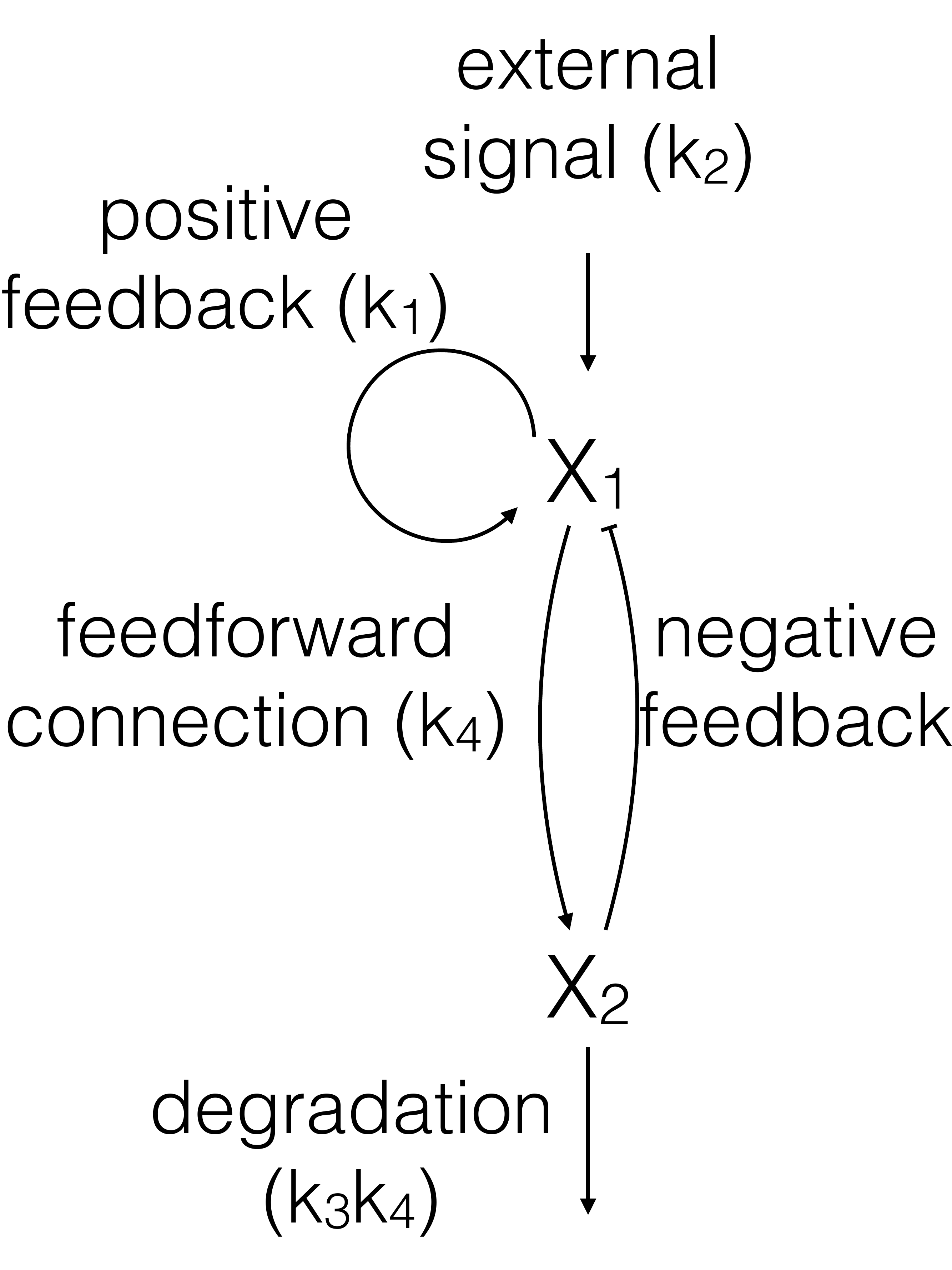}
\hskip 6.26mm
\raise 5.5cm
\hbox{(c)}
\hskip -10.2mm
\raise 5mm
\hbox{\includegraphics[width=.36\textwidth]{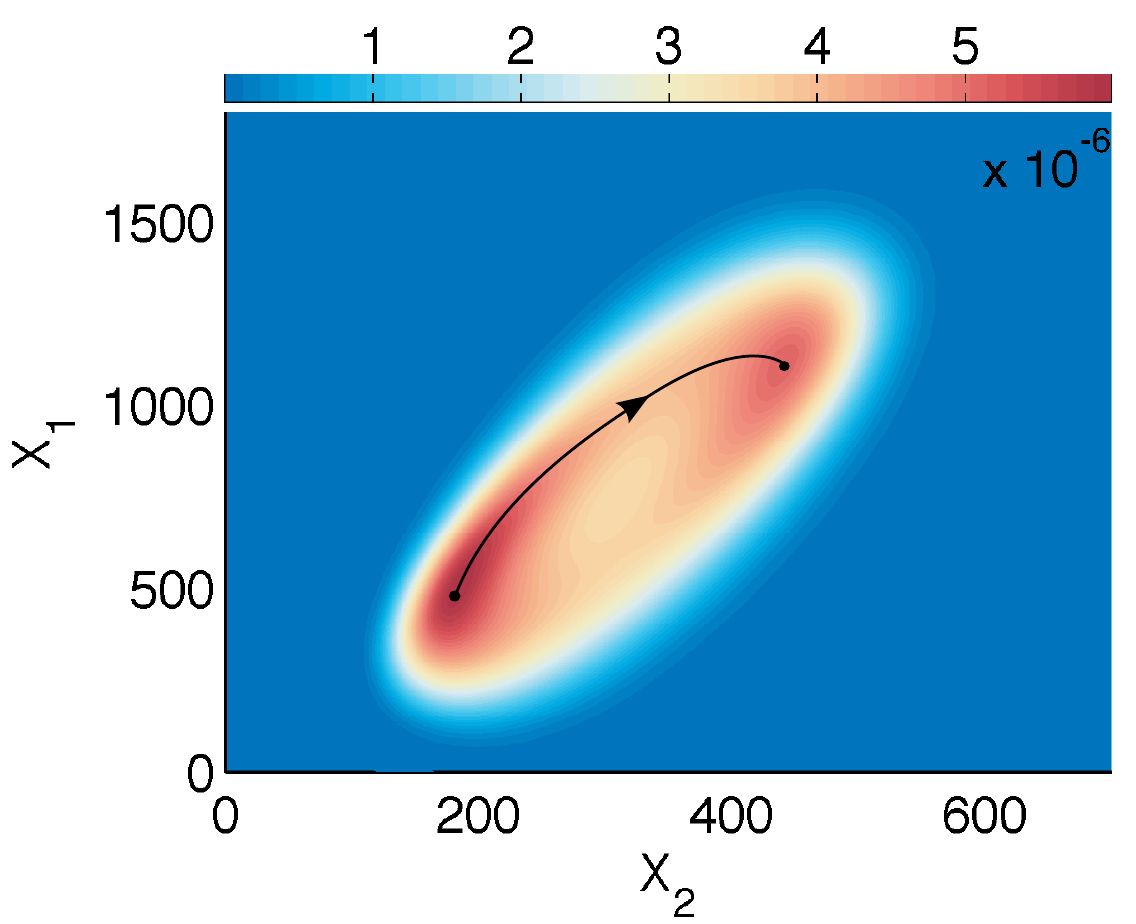}}
\hskip 2.26mm
\raise 5.5cm
\hbox{(d)}
\hskip -10.2mm
\raise 5mm
\hbox{\includegraphics[width=.36\textwidth]{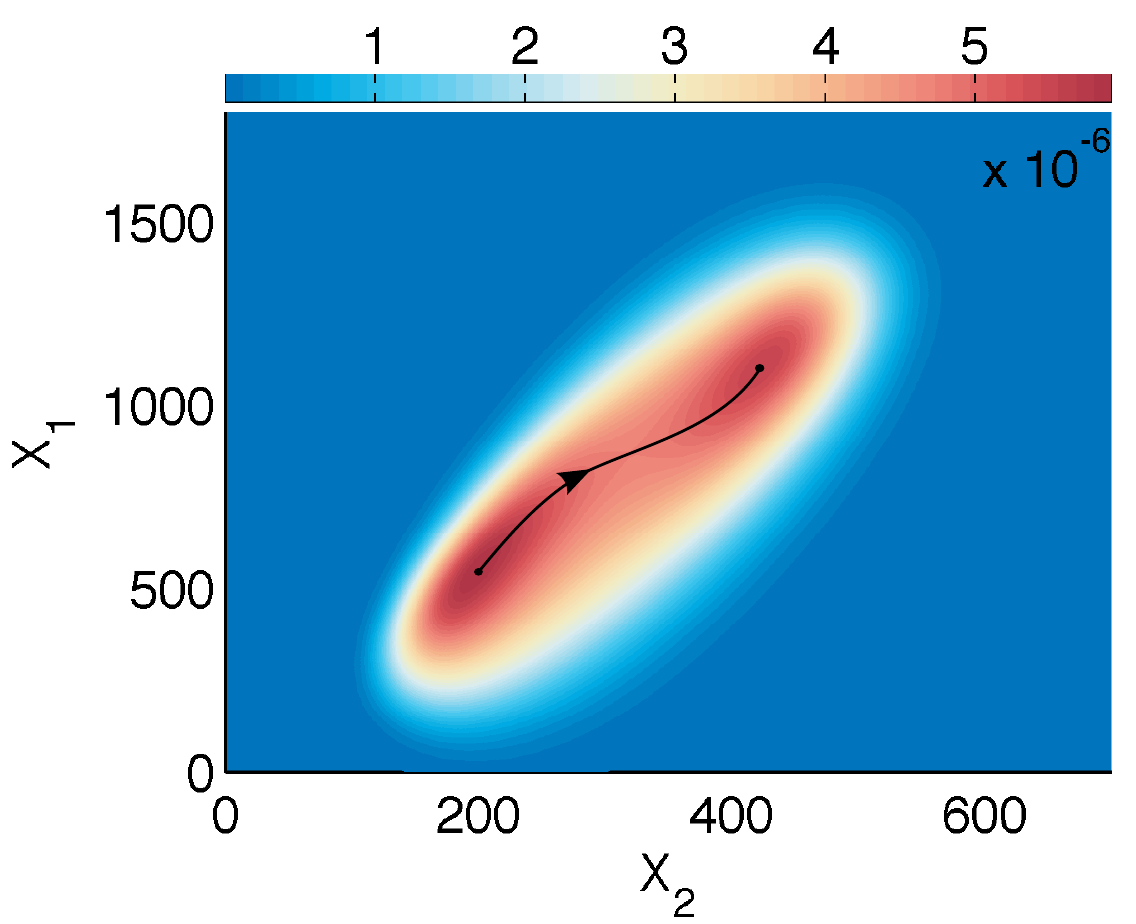}}
\hfill\break
\vskip -12.0mm
\noindent 
\raise 4.6cm
\hbox{(b)}
\hskip -8.5mm
\includegraphics[width=.31\textwidth]{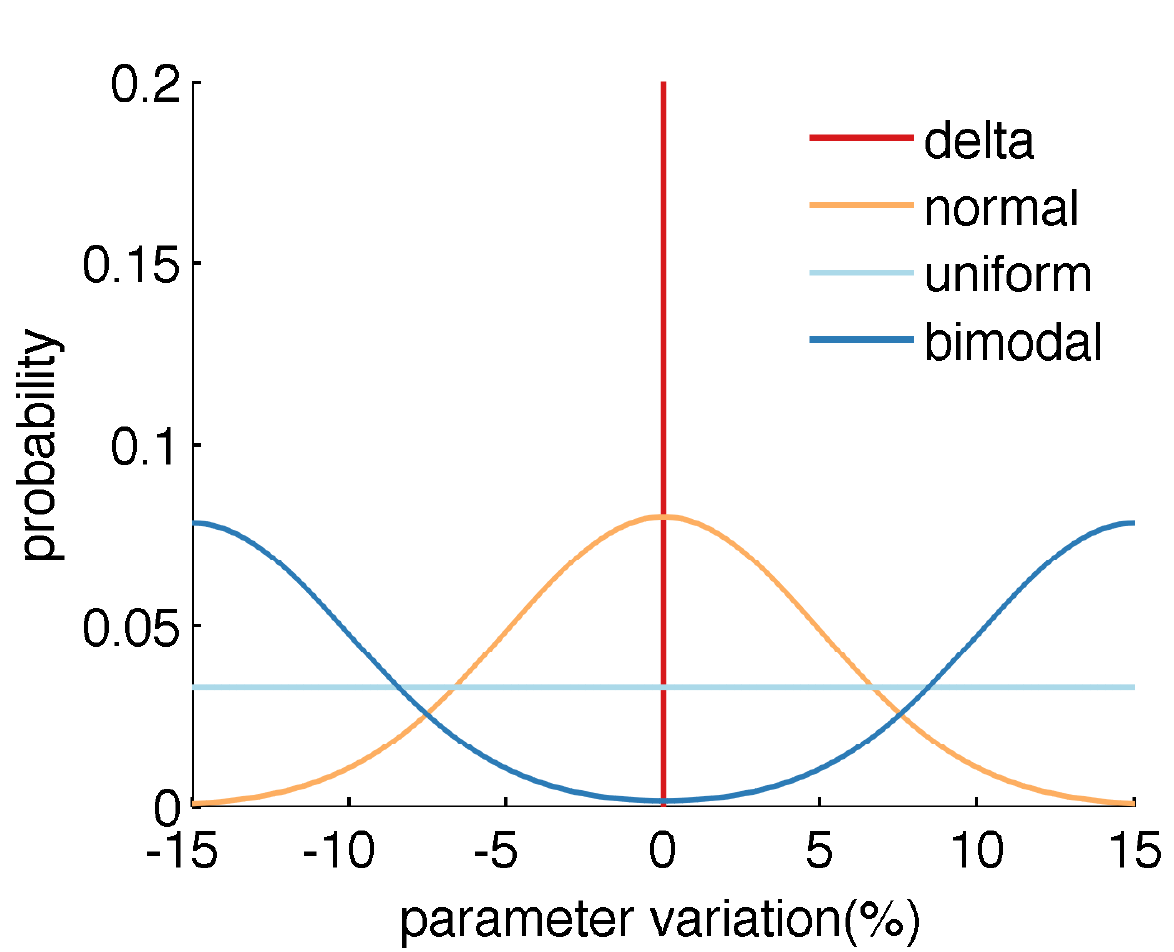} 
\hskip 3.4mm
\raise 4.6cm
\hbox{(e)}
\hskip -10.2mm
\raise -3mm
\hbox{\includegraphics[width=.36\textwidth]{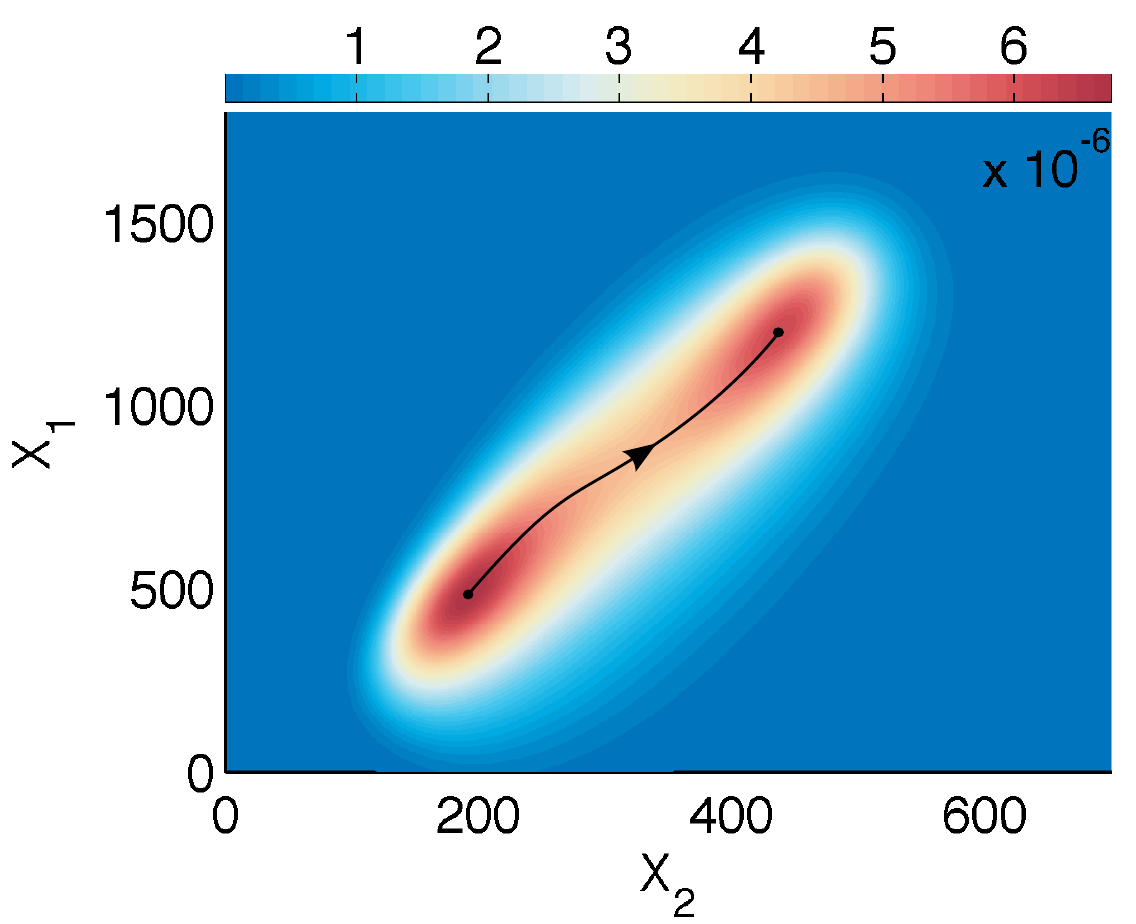}}
\hskip 2.26mm
\raise 4.6cm
\hbox{(f)}
\hskip -10.2mm
\raise -3mm
\hbox{\includegraphics[width=.36\textwidth]{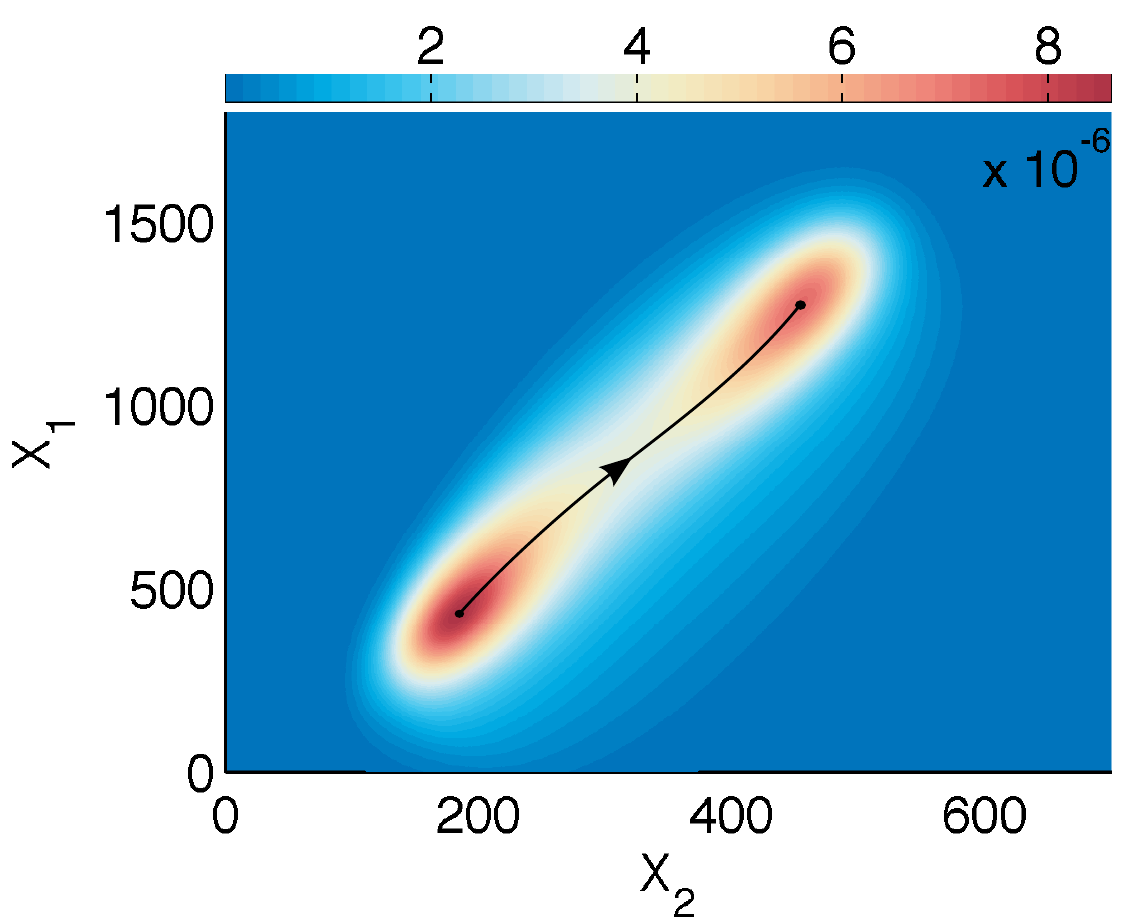}}
\caption{{\it 
Analysis of the FitzHugh-Nagumo system.}
(a)~{\it Schematic of the model.} \hfill\break
(b) {\it Four types of distributions of the extrinsic 
noise applied to model parameters.} \hfill\break
(c)--(f) {\it Steady state behaviour of the FitzHugh-Nagumo model
with} (c) {\it constant reaction
rates}, (d) {\it normal,} (e) {\it uniform,  and} (f)
{\it bimodal distribution of the model parameters. 
The tensor-structured computational
procedure follows formula} $(\ref{probdistindep})$.
{\it The directed black curves track the ridges of the stationary 
distribution, and depict the ``most-likely'' transition paths 
from the inhibited state to the excited state.}
\label{fig6}}
\end{figure}%

\subsection{Extrinsic noise in FitzHugh-Nagumo model}

We consider the effect of extrinsic fluctuations on an 
activator-inhibitor oscillator with simple negative feedback 
loop: the FitzHugh-Nagumo neuron model which is presented
in Figure~\ref{fig6}(a). Self-autocatalytic positive feedback 
loop activates the $X_1$ molecules, which are further triggered 
by the external signal. The species $X_2$ is enhanced by the 
feedforward connection and it acts as an inhibitor that turns 
off the signalling \cite{tsai2008robust}. 
{\modify
We perform robustness analysis based on the simulated tensor 
data in Section~\ref{sectab1} (summarized on the third line of
 Table~\ref{tabsimdetails}).}
In our computational examples, we assume that 
$q ( \mathbf{k} ) = q_1(k_1) \, q_2(k_1) \,
\ldots \, q_M(k_M)$, i.e. the invariant distributions
of rate constants $k_1$, $k_2$, \dots, $k_M$ are independent. 
Then (\ref{probdistint}) reads as follows
\begin{equation}
p(\mathbf{x}) = \int_{\Omegak} p ( \mathbf{x}|\mathbf{k}) 
\, q_1(k_1) \, \cdots \, q_M(k_M) \, \mathrm{d} \mathbf{k}.
\label{probdistindep}
\end{equation}
Extrinsic variability in the FitzHugh-Nagumo system
is studied in four prototypical cases of $q_i$, $i=1,2,\dots,M$:
(i) Dirac delta, (ii) normal, (iii) uniform, and (iv) bimodal distributions, 
as shown in Figure~\ref{fig6}(b). Since these distributions 
have zero mean, the extrinsic noise is not biased. 
We can then use this information about extrinsic noise to simulate the 
stationary probability distribution of intrinsic variables by
(\ref{probdistindep}).

When the extrinsic noise is omitted, the inhibited and excited 
states are linked by a volcano-shaped oscillatory probability 
distribution (Figure~\ref{fig6}(c)). At the 
inhibited state, $X_1$ molecules first get activated from the 
positive feedback loop, and then excite $X_2$ molecules by 
feedforward control. The 
delay between the excitability of the two molecular species gives rise 
to the path (solid line) describing switching from the inhibited state 
to the excited state (Figure~\ref{fig6}(c)). If the normal or
uniform noise are introduced to the extrinsic variables, then the path
becomes straighter (Figure~\ref{fig6}(d) and Figure~\ref{fig6}(e)). 
This suggests that, once $X_1$ molecules get excited or inhibited, $X_2$ 
molecules require less time to response.

GRNs with stronger negative feedback regulation 
gain higher potential to reduce the stochasticity. This argument 
has been both theoretically 
analysed~\cite{thattai2001intrinsic, swain2004efficient},
and experimentally tested for a plasmid-borne 
system~\cite{becskei2000engineering}. We have shown that 
the extrinsic noise reduces the delay caused by the feedback 
loop (Figure~\ref{fig6}(d)). If we further 
increase the variability of the extrinsic noise, then the delay
caused by the feedback loop is further reduced
(Figure~\ref{fig6}(e)). In the case of the bimodal distribution
of extrinsic fluctuations, the most-likely path linking the 
inhibited and excited states even shrinks into an almost straight 
line (Figure~\ref{fig6}(f)). This means 
that, for the same level of the inhibitor $X_2$, the number 
of the activator $X_1$ is lower, i.e.
the presented robustness analysis shows that the 
behaviour of stochastic GRNs with negative 
feedback regulation can benefit from the extrinsic noise.

{\modify
\section{Discussion}
}

{\modify

\noindent
We have presented the TPA of stochastic reaction networks
and illustrated that the TPA can (i) calculate and store the 
parametric steady state distributions; (ii) infer and analyse 
stochastic models of GRNs. To explore high dimensional state 
space $\Omega_\bold{x}$ and parameter space $\Omega_\bold{k}$, 
the TPA utilises a recently proposed low-parametric 
tensor-structured data format, as presented 
in equation (\ref{tensorrep}). Tensor methods have been recently
used to address the computational intensity of solving
the CME~\cite{dolgov2012,kazeev2013direct}. In this paper, 
we have extended these tensor-based approaches from solving the 
underlying equations to automated parametric analysis of the 
stochastic reaction networks. 
One notable advantage of the tensor approach lies in its ability 
to capture all probabilistic information of stochastic models 
all over the parameter space into one single tensor-formatted 
solution, in a way that allows linear scaling of basic operations 
with respect to the number of dimensions. Consequently, the 
existing algorithms commonly used in the deterministic framework 
can be directly utilized in stochastic models via the TPA.
In this way, we can improve our understanding of parameters 
in stochastic models.

To overcome technical (numerical) challenges, we have 
introduced two main approaches for successful computation 
of the steady state distribution. First, we compute it 
using the CFPE approximation which provides additional 
flexibility in discretising the state space $\Omegax$. The CFPE 
admits larger grid sizes for numerical simulations than 
the unit grid size of the CME. In this way, the resulting
discrete operator is better conditioned. We illustrate
this using a 20-dimensional problem introduced in the
last line of Table~\ref{tabsimdetails} and in \emph{SI Appendix} S2.4.
To compute the stationary distribution, a multilevel approach 
is implemented, where the steady state distribution is first 
approximated on a coarse grid, and then interpolated to a finer 
grid as the initial guess (see \emph{SI Appendix} S1.3 for more details).
The results are plotted in Figure~\ref{fig7}. 
Second, we introduce the adaptive inverse power iteration scheme 
tailored to current tensor solvers of linear systems, see 
\emph{SI Appendix} S1.3 for technical details. 
Since tensor linear solvers are 
less robust especially for ill-conditioned problems, it is necessary 
to carefully adapt the shift value during the inverse power iterations 
in order to balance the conditioning and sufficient speed of 
the convergence.

\begin{figure}[t]
\noindent
\rule{0pt}{0pt} \hskip 1mm
\raise 4.2cm \hbox{(a)}
\hskip -8mm
\includegraphics[scale=.43]{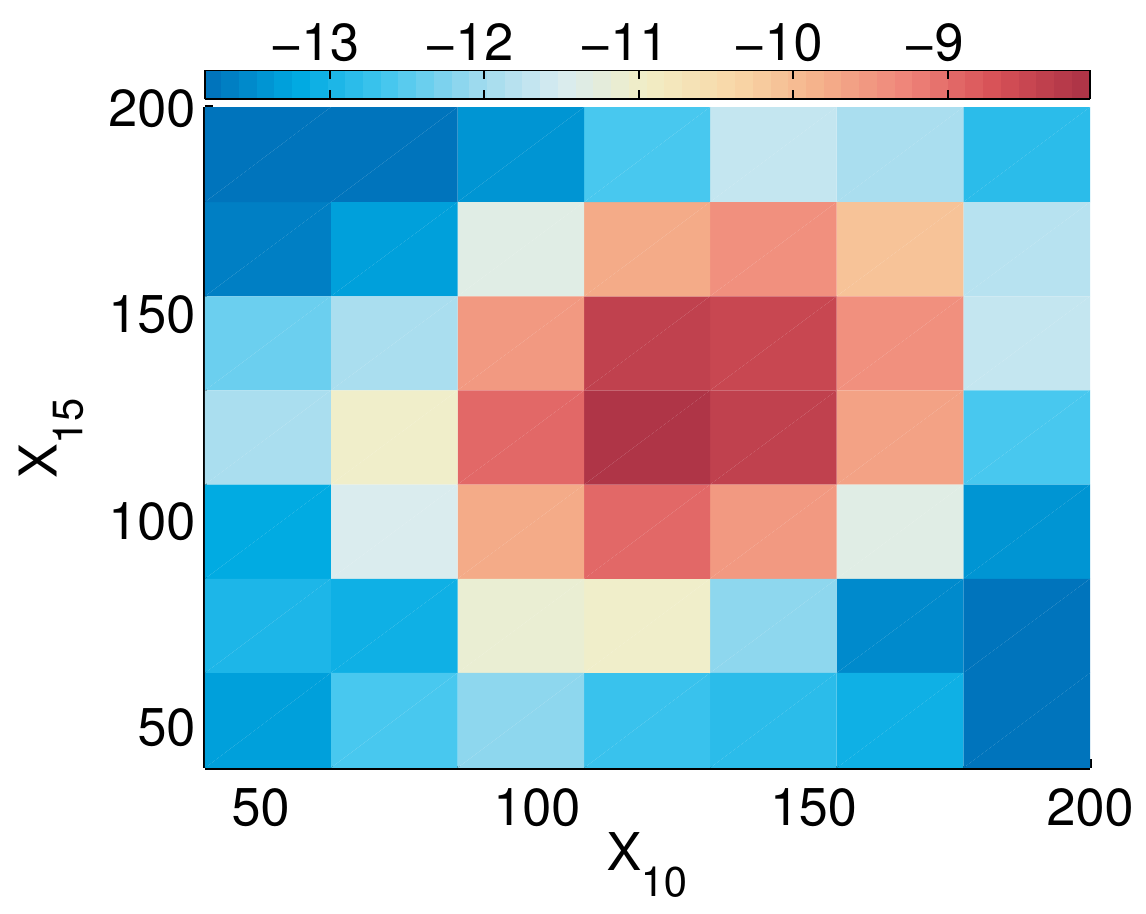}
\hskip 5mm
\raise 4.2cm \hbox{(b)}
\hskip -8mm
\includegraphics[scale=.43]{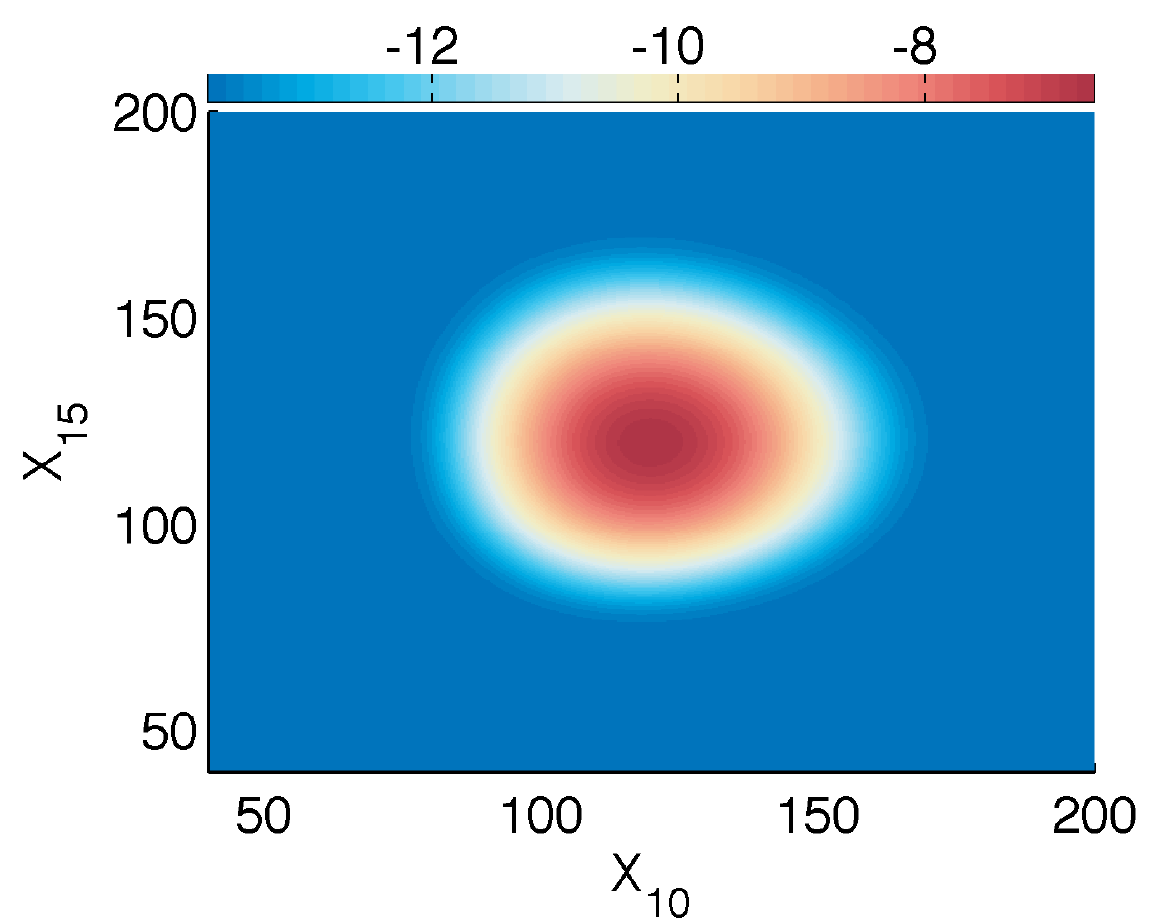}
\hskip 5mm
\raise 4.2cm \hbox{(c)}
\hskip -8mm
\includegraphics[scale=.43]{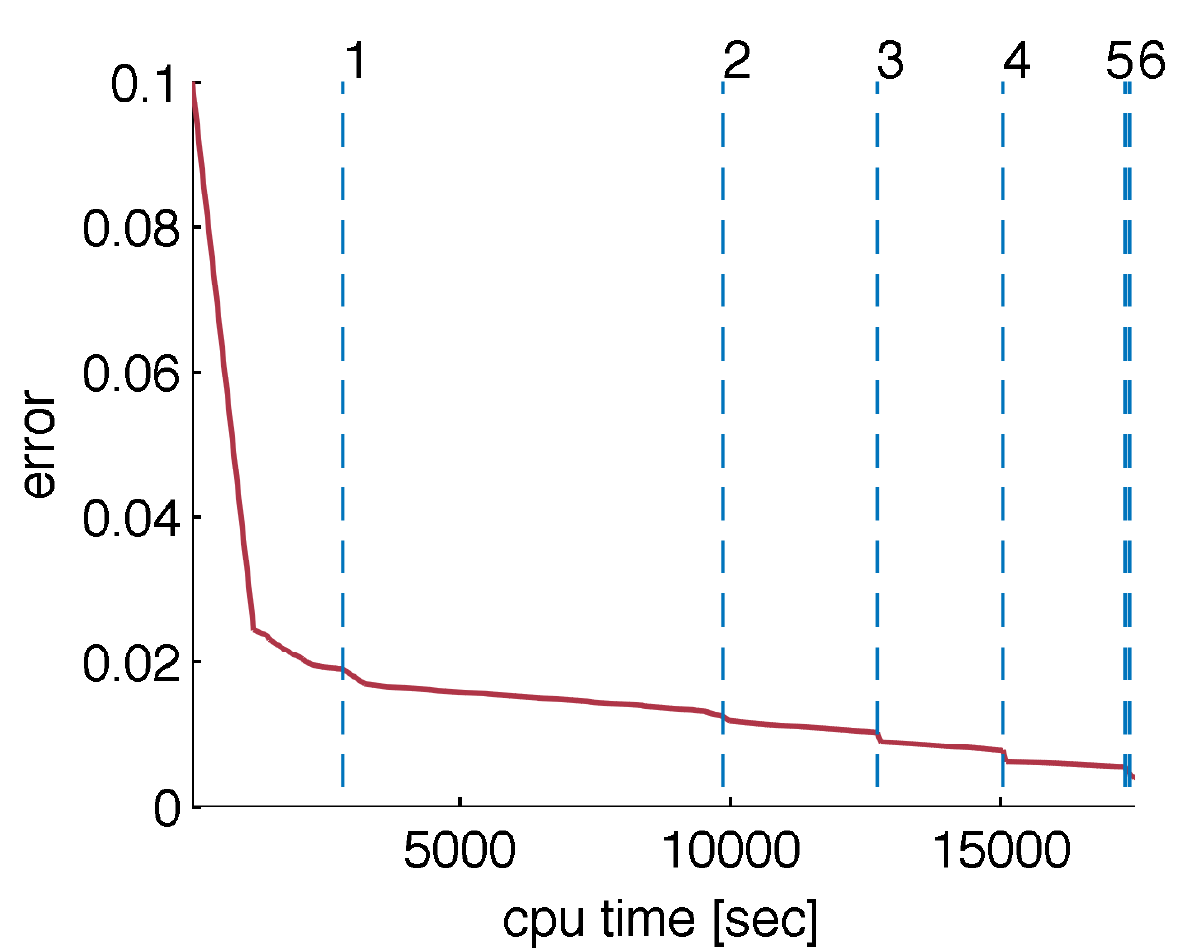}
\caption{{\it The computation of the stationary distribution
using the TPA for a $20$-dimensional reaction chain. 
The CFPE is successively solved on seven grid levels with
an increasing number of nodal points. The marginal stationary 
distribution in the $X_{10}$-$X_{15}$ plane computed on} 
(a) {\it the initial coarsest level;} \hfill\break
$\left.\right.$ \hskip 8.78cm (b) {\it the finest grid level.} \hfill\break
(c) {\it The convergence of the total error 
versus the computational time. The vertical dashed lines correspond 
to the grid levels. The grid size details on each grid level are given 
in SI Appendix} S2.4 {\it (Table} S8{\it ).}
\label{fig7}}
\end{figure}

Techniques for the parameter inference and bifurcation analysis of
stochastic models have been less studied in the literature
than the corresponding methods for the ODE models.
One of the reason for this is that the solution of the CME
is more difficult to obtain than solutions of mean-field ODEs.
This has been partially solved by  the widely-used Monte Carlo methods, 
such as the Gillespie SSA, which can be used to estimate 
the required quantities~\cite{Toni2009}. 
Advantages of Monte Carlo methods are especially their relative 
simplicity and easy parallelisation. The TPA provides an alternative 
approach. The TPA uses more complex data structures and algorithms
than the Gillespie SSA, but it enables to compute the whole probability
distribution for all combinations of parameter values at once. The TPA
stores this information in the tensor format. If the state and 
parameter spaces have higher number of dimensions, then the Monte
Carlo methods would have problems with storing computed stationary
distributions.
Another advantage of the TPA is that it produces smooth data, 
see e.g. Figure~\ref{fig3} for the data over the parameter space 
and Figure~\ref{fig5} for the data in the state space. This is 
important for a stable convergence in the gradient-based optimization
algorithms~\cite{McGill2012Efficient}, and for reliable analysis 
of stochastic bifurcations. Monte Carlo methods provide necessarily 
noisy and hence nonsmooth data that may cause problems for these methods.

Parameter inference of stochastic models can make use 
of various statistical measures, such as the variance and correlations.
Monte Carlo approaches are widely used to compute these quantities, 
but they may be computationally expensive. The TPA provides 
an alternative approach. Once we compute the stationary distribution 
for the desired ranges of parameter values and store it in the tensor 
format, we can use the tensor operation techniques (see 
\emph{SI Appendix} S1.4) 
to efficiently compute many different statistical measures 
from the same stationary distribution. If the results of the used 
statistical measure and chosen method are not satisfactory, we can 
modify or completely change both and try to infer the parameters again. 
Since the stationary distribution is stored, the modifications and 
changes can be done with low computational load. Namely, no stochastic
simulations are needed. In addition, since the stationary distribution 
contains complete information about the stochastic steady state, 
it can be used to compute practically any quantity for comparison with
experimental data. 
We have illustrated several different parametric studies in 
Figures~\ref{fig1}(b), \ref{fig2}, and \ref{fig3}. All these results are 
based
on a single tensor solution presented in Section~\ref{sectab1}
(see Table~\ref{tabsimdetails}). Let us also note that it
is relatively straightforward to use the TPA framework to study
the parameter sensitivity of stochastic systems (i.e. 
to quantify the dependence of certain quantities of interest on continuous
changes in model parameters). A systematic way for conducting the 
sensitivity analysis is illustrated in \emph{SI Appendix} S1.5, 
using the fission yeast cell cycle model.
}

\vskip 2mm

\paragraph*{Acknowledgments.}
The research leading to these results has received 
funding from the European Research Council under the European 
Community's Seventh Framework Programme  (FP7/2007-2013) /
ERC grant agreement No. 239870;
and from the People Programme (Marie Curie Actions) of the European Union's 
Seventh Framework Programme (FP7/2007-2013) under REA grant agreement 
no. 328008. Radek Erban would also like to thank 
the Royal Society for a University Research 
Fellowship; Brasenose College, University of Oxford, for a Nicholas 
Kurti Junior Fellowship; and the Leverhulme Trust for a Philip 
Leverhulme Prize.
Tom\'a\v{s} Vejchodsk\'y also acknowledges the support of RVO: 67985840.

\newpage

\appendix

% For section headers starting with S
\renewcommand{\thesection}{S\arabic{section}}
\renewcommand{\thesubsection}{\thesection.\arabic{subsection}}
 
% Making SOM Equations e.g. (S1), (S2), etc
% Requires AMS
\makeatletter %% With ams
\def\tagform@#1{\maketag@@@{(S\ignorespaces#1\unskip\@@italiccorr)}}
\makeatother
 
% figures, e.g. Figure S1:
\makeatletter
\makeatletter \renewcommand{\fnum@figure}
{\figurename~S\thefigure}
\makeatother

% tables, e.g. Table S1:

\makeatletter
\makeatletter \renewcommand{\fnum@table}
{\tablename~S\thetable}
\makeatother
 
\newcommand{\citenumfont}[1]{\textit{S#1}}

\setcounter{equation}{0}

\title{\noindent \Large \textmd{Supporting Information Appendix for:} \\
\vskip 2mm
\noindent \LARGE \emph{Tensor methods for parameter estimation and bifurcation 
analysis of stochastic reaction networks}
\\ \vskip 1mm}

\noindent 
\author{
\large
Shuohao Liao\footnote{Mathematical Institute, University of Oxford, 
Radcliffe Observatory Quarter, Woodstock Road, Oxford OX2 6GG, 
United Kingdom; e-mails: liao@maths.ox.ac.uk; erban@maths.ox.ac.uk}, 
Tom\'{a}\v{s} Vejchodsk\'{y}\footnote{Institute of Mathematics, 
Czech Academy of Sciences, Zitna 25, 
115 67 Praha 1, Czech Republic; e-mail vejchod@math.cas.cz}, 
Radek Erban$^{\ast}$
}

\date{}

\maketitle

\section{Methods}

\label{secmethods}

All models studied by the TPA are given in terms of well-mixed 
chemical systems 
where the system state changes according to the chemical reactions 
(1) (here, equation label (1) refers to the corresponding
equation in the paper). %\eqref{eq: mass-action reaction}. 
Probability that a reaction occurs is determined by the propensity function
\begin{equation}   
\label{eq: propensity}
\alpha_j(\mathbf{x},k_j) 
= k_j \, \tilde{\alpha}_j(\mathbf{x}),
\quad j=1,2,\dots,M,
\end{equation}
with the non-parametric part $\tilde{\alpha}_j(\mathbf{x})$ given by
$$
\tilde{\alpha}_j(\mathbf{x}) = 
\exp \left[ \left(
1 - \sum_{i=1}^N \nu_{j,i}^-
\right)
\log V \right] 
\prod_{i=1}^N 
(\nu_{j,i}^-) !
\binom{x_i}{\nu_{j,i}^-},
$$
where $V$ is the volume of the reactor.
The stationary distribution $p \big( \mathbf{x}\,|\,\mathbf{k} \, \big)$, where
$\mathbf{k}=(k_1,k_2,\ldots,k_K)^T$ is the subset of kinetic rate
constants for which the parameter analysis is considered,
can be computed as the exact solution of the chemical 
master equation~\cite{gillespie1992rigorous}.
However, for the computational reasons, we will approximate it by the 
solution of the stationary chemical Fokker-Planck 
equation~[18], which can be written as
$$ 
\mathcal{A}(\mathbf{x},\mathbf{k}) \,
p \big( \mathbf{x}\,|\,\mathbf{k} \, \big)
= 0,
$$ 
where
\begin{eqnarray}   
\nonumber
\mathcal{A}(\mathbf{x},\mathbf{k})
\,
p \big( \mathbf{x}\,|\,\mathbf{k} \, \big)   
&=& 
- \sum_{i=1}^N \frac{\partial}{\partial x_i} 
\left( \sum_{j=1}^M \nu_{j,i}
\,
\alpha_j (\mathbf{x},k_j) 
\,
p \big( \mathbf{x}\,|\,\mathbf{k} \, \big) \right) \\
&+&
\frac{1}{2} \sum_{i,i^{\prime} = 1 }^N 
\frac{\partial^2}{\partial x_i \partial x_{i^{\prime}}} 
\left( \sum_{j=1}^M 
\nu_{j,i} 
\,
\nu_{j,i^{\prime}} 
\,
\alpha_j (\mathbf{x},k_j) 
\,
p \big( \mathbf{x}\,|\,\mathbf{k} \, \big) \right),
\label{eq: CFPoperator}
\end{eqnarray}
is the parametric Fokker-Planck operator. We use the tensor structures 
to compute $p \big( \mathbf{x}\,|\,\mathbf{k} \, \big)$ 
simultaneously for ranges of values of reaction rates
$\mathbf{k}$. To achieve this, we split the model parameters from the state 
variables in a multiplicative way. Considering the definition of 
propensity functions \eqref{eq: propensity}, we can split the
parametric Fokker-Planck operator \eqref{eq: CFPoperator} into $M$ terms as
\begin{equation} 
\label{eq: split Fokker-Planck operator}
\mathcal{A}(\mathbf{x},\mathbf{k}) 
= 
k_1 \mathcal{A}^{[1]}(\mathbf{x}) + \cdots + k_M \mathcal{A}^{[M]}(\mathbf{x}),
\end{equation}
where the non-parametric operator $\mathcal{A}^{[j]}(\mathbf{x})$ 
describes the normalised transition properties of the $j$-th reaction, 
and is defined (for any twice differentiable function $f$) by
\begin{equation}   
\label{eq: non-parametric operator}
\mathcal{A}^{[j]}(\mathbf{x}) 
\, f(\mathbf{x})
= 
- \sum_{i=1}^N  \nu_{j,i} \,
\frac{\partial}{\partial x_i}  
\Big( \tilde{\alpha}_j(\mathbf{x}) \, f(\mathbf{x}) \Big) 
+ 
\frac{1}{2} \sum_{i,i^{\prime} = 1 }^N 
\, \nu_{j,i^{\prime}} \, \nu_{j,i} 
\frac{\partial^2}{\partial x_i \partial x_{i^{\prime}}}
\Big( \tilde{\alpha}_j(\mathbf{x}) \, f(\mathbf{x}) \Big).
\end{equation}
Let us note that the definition of propensity functions 
\eqref{eq: propensity} relies on the law of mass action. 
However, the TPA methodology is applicable even for more 
general definitions~\cite{wu2011constructing}. If the 
propensity functions depend nonlinearly on the kinetic rates, 
as in~\cite{barkai2000biological} for example, 
then the TPA methodology can be used provided a multiplicative 
splitting \eqref{eq: split Fokker-Planck operator} of the parametric 
Fokker-Planck operator is possible. Such splitting is always possible 
if the propensities can be written as a product of two terms, where 
the first term depends only on kinetic rates, and the second 
term on state variables. 

The splitting of the parametric Fokker-Planck 
operator~\eqref{eq: split Fokker-Planck operator} implies a 
perfect collinear relationship between the kinetic rate parameters. 
In the context of parameter estimation this means that a single 
constraint (like mean and variance) restricts the original 
$K$-dimensional parameter space to an $(K-1)$-dimensional 
subspace of parameter values that comply with the constraint.
Therefore, if a direct comparison between a model and a sample is not 
possible, a necessary condition to statistically infer $K \le M$ 
parameters of a stochastic system is to define at least $K$ constraints.

\subsection{Tensorization}
\label{sectens}

We consider the state variable $\mathbf{x}$ in a bounded 
domain $\Omegax \subset (0,\infty)^N$. Similarly, the kinetic 
rates $\mathbf{k}$, which are varied during the parametric analysis,
are considered in $\Omegak \subset [0,\infty)^K$ where $K \le M$. 
In order to utilize the tensor structures, we assume 
that $\Omegax = \cI_1 \times \cdots \times \cI_N$ and 
$\Omegak = \cJ_1 \times \cdots \times \cJ_K$, where 
$\cI_d = (a^x_d,b^x_d)$, $d=1,2,\dots,N,$ are open intervals 
and $\cJ_\ell = [a^k_\ell,b^k_\ell]$, $l=1,2,\dots,K,$
are closed intervals. 

{\modify
We consider homogeneous Dirichlet boundary conditions on the 
boundary of $\Omega_\mathbf{x}$. We approximate the stationary 
distribution by the (normalized) eigenfunction of 
$\mathcal{A}(\mathbf{x},\mathbf{k})$ corresponding to the eigenvalue 
closest to zero. Since the Fokker-Planck operator 
$\mathcal{A}(\mathbf{x},\mathbf{k})$
is an elliptic operator, the largest eigenvalue converges to 
0 from below as the size of $\Omega_\mathbf{x}$ increases 
to infinity. In particular, if we choose a sufficiently large
computational domain $\Omega_\mathbf{x}$, then the largest 
eigenvalue
will be close to zero and the Dirichlet boundary conditions 
will not cause any substantial error.}

The chemical Fokker-Planck operator \eqref{eq: non-parametric operator}
is discretized in $\Omegax$ by the finite difference method.
We consider tensor grids \cite{nobile2008sparse} in both $\Omegax$ 
and $\Omegak$.
The tensor grid in $\Omegax$ has nodes $(x_{1,i_1}, \dots, x_{N,i_N})$, 
$i_d = 1,2, \dots,n_d$, $d=1,2,\dots,N$. There are $n_d$ points 
$x_{d,i_d} = a^x_d + i_d h^x_d$, $i_d = 1,2, \dots,n_d$, in every 
$\cI_d$ with the grid size $h^x_d = (b^x_d-a^x_d)/(n_d+1)$, 
$d=1,2,\dots,N$. Similarly, we define tensor grid 
$(k_{1,j_1}, \dots, k_{K,j_K})$ in $\Omegak$,
where $k_{\ell,j_\ell} = a^k_\ell + (j_\ell-1) h^k_\ell$, 
$j_\ell=1,2,\dots,m_\ell$, form a uniform partition of 
$\cJ_\ell$ with the grid size $h^k_\ell = (b^k_\ell - a^k_\ell)/(m_\ell-1)$, 
$\ell=1,2,\dots,K$. Note that the boundary points $a^x_d$ and $b^x_d$, 
$d=1,2,\dots,N$, are not present in the tensor grid due to the  
Dirichlet boundary conditions.

The values of the stationary distribution 
$p \big( \mathbf{x}\,|\,\mathbf{k} \, \big)$ at 
the nodal points are organized as an $(N+K)$-dimensional tensor 
$\tenp \in \R^{n_1\times \cdots \times n_N \times m_1 \times \dots \times m_K}$
with entries
\begin{equation}
\label{eq:tensorp}
\tenp_{i_1, \dots, i_N, j_1\dots,j_K}
 = 
p(x_{1,i_1}, \dots, x_{N,i_N}|k_{1,j_1}, \dots, k_{K,j_K}).
\end{equation}

In the traditional matrix-vector approach, we would organize the 
entries of $\tenp$ into a long vector. However, the tensor
structure is more natural, because it corresponds to the original 
physical position of the nodes within the state and parameter space 
\cite{donoho2000high}.
Finally, let us note that if $n=n_1=\cdots=n_N$ and $m=m_1=\cdots=m_K$ 
then there is $n^N m^K$ entries in the tensor $\tenp$. Thus,
the number of memory places 
to store the tensor $\tenp$ grows exponentially with $N$ and $K$.
In the next subsection, we present the main idea of the separated 
representation of tensors that allows to solve this problem.

\subsection{Separation of dimensions}

The main idea of the separated (or low-parametric) representation 
is to approximate tensor $\tenp$ by the following sum 
of rank-one tensors:
\begin{equation} 
\label{eq: solution: discrete separated expansion}
\tenp \approx \sum_{r=1}^R
\underset{\mathrm{state\ space}} 
{ \underbrace{ \mathbf{\phi}^{[r]}_{1} \otimes \cdots \otimes \mathbf{\phi}^{[r]}_{N}}}
\otimes
\underset{\mathrm{parameter\ space}}{ \underbrace{ \mathbf{\psi}^{[r]}_1 
\otimes \cdots \otimes \mathbf{\psi}^{[r]}_K } },
\end{equation}
where $\mathbf{\phi}^{[r]}_{d} \in \R^{n_d}$, $d=1,2,\dots,N,$
and $\mathbf{\psi}^{[r]}_\ell \in \R^{m_\ell}$, $\ell=1,2,\dots,K,$ 
are factor vectors, $R$ is known as the separation rank, and symbol 
$\otimes$ denotes the tensor product of vectors \cite{lynch1964direct}. 
Let us recall that the tensor product 
$\mathbf{v}_1 \otimes \mathbf{v}_2 \otimes 
\cdots \otimes \mathbf{v}_N$ of vectors 
$\mathbf{v}_d \in \R^{n_d}$, $d=1,2,\dots,N$, is defined as a 
tensor $\tenv \in \R^{n_1\times n_2\times\cdots\times n_N}$
with entries 
$v_{i_1,i_2,\dots,i_N} = v_{1,i_1} \, v_{2,i_2} \cdots \, v_{N,i_N}$.

Representation \eqref{eq: solution: discrete separated expansion} 
has the potential to solve high-dimensional problems. Indeed, if 
we consider for simplicity $n=n_1 = \cdots = n_N$ and $m=m_1 = \cdots = m_K$ 
then the representation \eqref{eq: solution: discrete separated expansion} 
requires to store $(nN+mK)R$ numbers only. For moderate values of $R$ this 
is substantially less than the number of entries of $\tenp$.
Moreover, low-parametric representations such 
as \eqref{eq: solution: discrete separated expansion} enable to 
perform algebraic operations in an efficient way,
see Section~\ref{subsectensoroperations}.

The accuracy of the separated representation 
\eqref{eq: solution: discrete separated expansion} depends on the choice 
of the factor vectors and on the size of the tensor rank $R$. Clearly, the 
higher rank enables higher accuracy, but requires higher computational and 
storage costs. In practical computations, the rank $R$ is dynamically 
controlled using algorithms for tensor truncation, 
see Section~\ref{subsecsolveCFPE}. Let us note that the representation 
\eqref{eq: solution: discrete separated expansion} is known as the 
canonical polyadic decomposition~\cite{kruskal1977three}. However, 
due to reasons connected with the stability of the tensor truncation 
algorithms, it is not suitable for actual computation and more stable 
tensor formats have to be employed~[22]. We have 
introduced the canonical polyadic decomposition 
\eqref{eq: solution: discrete separated expansion} due to its 
simplicity to illustrate the main idea of the separate representation of tensors.

For certain simple problems, like birth-death process, the separable 
representation of the stationary distribution can be 
derived explicitly. However, in general, we have to compute the stationary 
distribution in the form \eqref{eq: solution: discrete separated expansion}.
To achieve this, we need to express the discretized Fokker-Planck operator 
in a separable form as well. Based on the structure of 
$\mathcal{A}(\mathbf{x}|\mathbf{k})$ in 
\eqref{eq: split Fokker-Planck operator}, 
the discretization of the parametric Fokker-Planck operator can be divided 
into two steps: decomposing the non-parametric part 
(see Section \ref{secdnp}) and the parametric part
(see Section \ref{secdpp}).

\subsubsection{Decomposition of the non-parametric part}
\label{secdnp}

We use the finite differences to discretize the derivatives in the 
non-parametric operators $\mathcal{A}^{[j]}(\mathbf{x})$ in 
\eqref{eq: non-parametric operator}, see e.g \cite{smith1985numerical}. 
The separated tensor representation does not require 
high-dimensional difference stencils. Instead, just one-dimensional 
differences are needed. Further, since the standard finite difference 
discretizations of differential operators yield matrices, we organize 
their entries naturally into tensors. In this situation we speak about 
tensor matrices and denote them in capital bold font. The idea is exactly 
the same as in \eqref{eq:tensorp}, where we organized a long vector 
into a tensor.

Thus, the finite difference matrix approximating the non-parametric 
operator $\mathcal{A}^{[j]}(\mathbf{x})$ in \eqref{eq: non-parametric operator} 
can be expressed as the following tensor matrix:
\begin{equation} 
\label{eq: tensorised non-parametric}
\mathbf{A}^{[j]} = - \sum_{i =1}^N \nu_{j,i} \mathbf{G}^{[i;j]}
+ \frac{1}{2} \sum_{i,i' =1}^N \nu_{j,i} \nu_{j,i'} 
\mathbf{F}^{[i,i';j]},
\quad j=1,2,\dots,M,
\end{equation}
where tensor matrices $\mathbf{G}^{[i;j]}$ and $\mathbf{F}^{[i,i';j]}$ 
refer to tensor-structured discretizations of the summands in the first and 
second sums in \eqref{eq: non-parametric operator}, respectively, 
and are determined by
\begin{align*}
	\mathbf{G}^{[i;j]} =& \tilde{v}_j \, H^{[j]}_{1} \otimes 
	\cdots \otimes 
	D_i H^{[j]}_{i} \otimes \cdots \otimes H^{[j]}_{N}, \\
	\mathbf{F}^{[i,i';j]} =& \tilde{v}_j \, H^{[j]}_{1} \otimes 
	\cdots \otimes 
	D_i H^{[j]}_{i}  \otimes  \cdots \otimes 
	D_{i'} H^{[j]}_{i'} \otimes \cdots 
	\otimes H_N^{[j]}, \qquad \mbox{for} \quad i < i^\prime, \\
	\mathbf{F}^{[i,i;j]} =& \tilde{v}_j \, H^{[j]}_{1} \otimes 
	\cdots \otimes 
	D_i D_i H^{[j]}_{i} \otimes \cdots 
	\otimes H_N^{[j]},
\end{align*}
where the volume scaling coefficient is
$\tilde{v}_j = \exp\left[(1 - \sum_{i=1}^N \nu_{j,i}^-)\log V \right]$.
Here, $H^{[j]}_i \in \R^{n_i\times n_i}$ and $D_i \in \R^{n_i\times n_i}$ 
for $i=1,\ldots,N$ and $j=1,\ldots,M$ are matrices and, thus, the 
tensor product $\otimes$ works in the same way as the Kronecker product.
Matrix $D_i$ is the central difference matrix with entries 
$-1/(2h^x_i)$ 
and $1/(2h^x_i)$ distributed along its super- and sub-diagonal, 
respectively. 
Matrix $H^{[j]}_i$ is diagonal with diagonal entries
$$
H^{[j]}_i(\ell,\ell) = (\nu_{j,i}^-)!\binom{x_{i,\ell}}{\nu_{j,i}^-}
\qquad \mbox{for} \;\;\; \ell=1,2,\ldots, n_i.
$$
We observe that tensor matrices $\mathbf{G}^{[i;j]}$ and 
$\mathbf{F}^{[i,i';j]}$ are expressed in a separated representation 
similar to \eqref{eq: solution: discrete separated expansion} with the 
separation rank $R=1$. Consequently, the non-parametric operator 
$\mathbf{A}^{[j]}$ in \eqref{eq: tensorised non-parametric} 
admits separable representation of rank $R = N(N+1)/2+N = N^2/2 + 3N/2$. 
Thus, any further algebraic operation on $\mathbf{A}^{[j]}$ would 
contribute to the overall complexity growing quadratically in terms 
of number of chemical species.

\subsubsection{Decomposition of the parametric part}
\label{secdpp}

Having the low-parametric discrete tensor-structured representations 
\eqref{eq: tensorised non-parametric} of the non-parametric operators 
$\mathbf{A}^{[j]}$, we write a discrete tensor-structured 
representation of the parametric Fokker-Planck operator 
\eqref{eq: split Fokker-Planck operator} as
\begin{equation} 
\label{eq: sylverster matrix}
\mathbf{A} 
=  
\mathbf{A}^{[1]} \otimes K_1 \otimes I_2 \otimes \cdots 
\otimes I_M +\mathbf{A}^{[2]} \otimes I_1 \otimes K_2 \otimes \cdots 
\otimes  I_M  + \cdots + \mathbf{A}^{[M]} \otimes I_1 \otimes I_2 
\otimes \cdots \otimes K_M,
\end{equation}
where $K_j\in\R^{m_j\times m_j}$ denotes a diagonal matrix whose 
diagonal entries correspond to the grid nodes of the $j$-th parameter, 
i.e., $K_j(\ell,\ell) = k_{j,\ell}$ for $\ell=1,2,\ldots,m_j$ and $j=1,\dots,M$.

Equation \eqref{eq: sylverster matrix} is a low-parametric 
tensor representation of the discretized parametric Fokker-Planck 
operator with separation rank $M(N^2/2 + 3N/2)$. This rank grows 
linearly with the number of chemical reactions $M$ and quadratically 
with the number of chemical species $N$. Then, the parametric steady 
state distribution of the form \eqref{eq: solution: discrete separated expansion} 
is solved as the eigenvector of $\mathbf{A}$ corresponding to the 
eigenvalue closest to zero (see Section \ref{subsecsolveCFPE}).

\subsection{Solving the stationary CFPE in tensor format}
\label{subsecsolveCFPE}

Let $\mathbf{A}$ be the tensor-structured parametric Fokker-Planck 
operator assembled in \eqref{eq: sylverster matrix}. Our goal
is to approximate the stationary distribution by the eigenvector $\tenp$ 
corresponding
to the eigenvalue $\lambda_{\textrm{min}}$ which is closest to zero, i.e.
\begin{equation}
\label{eq: stationary equation in tensor format}
\mathbf{A} \tenp = \lambda_{\textrm{min}} \tenp.
\end{equation}
A standard method to find the required eigenpair of $\mathbf{A}$ is 
the inverse power method, and here we modify the original algorithm 
for better implementations in tensor-structured computations.

\paragraph{Adaptive inverse power algorithm.}
The main building block is the fact that, beginning with an initial guess 
$\tenp_0$ and given a shift value $\sigma$, the inverse power scheme,
\begin{equation} 
\label{eq: inverse iteration}
(\mathbf{A} - \sigma \mathbf{I}) \tenp_{k+1} =
\frac{\tenp_k}{\|\tenp_k\|}, \quad k=0, 1, \ldots
\end{equation}
would converge to the eigenvector corresponding to the eigenvalue closest to 
the chosen shift $\sigma$, provided that the eigenvalue is of multiplicity
one. Since all eigenvalues of the Fokker-Planck operator have negative real 
parts, we choose $\sigma \ge 0$. We do it adaptively based on the performance
of the tensor linear solver, i.e. 
$\sigma \equiv \sigma_k$ in~(\ref{eq: inverse iteration}).

We apply the alternating minimum energy method (AMEN) 
\cite{Dolgov2013Alternating} to solve the linear 
system \eqref{eq: inverse iteration}. Given an initial $N$-dimensional 
tensor $\tenp$, the AMEN method minimises the residual in
a single dimension at a time with other dimensions fixed, and alternates the 
dimensions from 1 to $N$. The entire \emph{sweep} repeats until a 
convergence criterion is satisfied. Typically, smaller shift $\sigma$ makes 
the whole inverse power method converge faster to the steady state solution, 
however, within each inverse
iteration \eqref{eq: inverse iteration}, the AMEN may require many 
sweeps to achieve 
a reasonable tolerance. Thus, our strategy is to double the shift  
value $\sigma$ when the solver reach certain upper threshold, 
and half $\sigma$ to seek for better convergence for the whole 
procedure when the AMEN converges with only a few sweeps.

Another extension arises from a feature of tensor-structured data format. 
The tensor separation rank $R$ can increase rapidly over successive algebraic 
operations, making the representation untenable. To avoid uncontrollable 
growth of
the separation rank throughout the computation, we need to reduce it by 
adaptively 
changing the involving factor vectors while maintaining the required accuracy. 
This
procedure is usually called tensor truncation:
\begin{equation} 
\label{eq: tensor truncation}
\tenp^* = \Gamma (\tenp),
\end{equation}
where operator $\Gamma$ is the truncation operator, and 
$\mathrm{rank}(\tenp^*) < \mathrm{rank}(\tenp)$.
Although finding the optimal tensor separation rank is still an open 
question of the 
ongoing research, the tensor train format, together with its SVD-based 
tensor 
truncation algorithm, is a stable and useful prototype for our 
implementations,
and we refer the readers to~[22] for further details.

Consequently, the adaptive inverse power method used in
the TPA is summarised as follows:
\begin{itemize}
\item[]
\begin{itemize}
\item[{\bf Step 0.}] 
Initialize: initial guess $\tenp_0$; shift value $\sigma = \sigma_0$; 
stopping criteria $\varepsilon$; maximum number of AMEN sweeps in each inverse
iteration $N_{max}$; thresholds to increase ($N_{in}$)
and decrease ($N_{de}$) the shift value.
\item[{\bf Step 1.}] 
Solve the $k$-th tensor-structured inverse iteration 
\eqref{eq: inverse iteration} up 
to $N_{max}$ sweeps.
\item[{\bf Step 2.}] 
Check the number of sweep $N_{comp}$ for the AMEN solver to converge:
\begin{itemize}
	\item[{\bf 2a.}] 
	If $N_{comp} > N_{in}$, let $\sigma = 2\sigma$ and jump back to Step 1.
	\item[{\bf 2b.}] 
	If $N_{de} < N_{comp} \leq N_{in}$, go to Step 3.
	\item[{\bf 2c.}] 
	If $N_{comp} \leq N_{de}$, $\sigma = \sigma / 2$ and go to Step 3.
\end{itemize}
\item[{\bf Step 3.}] 
Truncate the tensor separation rank as in \eqref{eq: tensor truncation}.
\item[{\bf Step 4.}] 
Check the stopping criteria:
\begin{itemize}
	\item[{\bf 4a.}]
	 If $\| \tenp_{k+1} - \tenp_{k} \| > \varepsilon$,
	 let $\tenp_{k+1} = \tenp_{k}$ and  $k = k+1$,
	 and jump to Step 1.
	\item[{\bf 4b.}] 
	If $\| \tenp_{k+1} - \tenp_{k} \| \leq \varepsilon$,
	return $\tenp_{k+1}$ and exit.
\end{itemize}
\end{itemize}
\end{itemize}

\paragraph{Multi-level acceleration.}
When the dimensionality of the problem is large, the adaptive scheme 
discussed above may converge slowly, because
on a fixed grid size, the AMEN requires very large shift value $\sigma$ 
to solve 
\eqref{eq: inverse iteration}. Thus, the TPA makes use of a multilevel 
scheme to accelerate the solution process for high-dimensional problems. 
The system \eqref{eq: stationary equation in tensor format} is first 
solved on a coarse grid with grid size $2h$. The approximated stationary 
solution is then interpolated to a fine grid with grid size $h$ and used
as an initial guess. The method continues to solve the system on 
finer grids until some convergence criteria are achieved.
	
A key step in the multilevel approach is the interpolation, or 
prolongation, matrix that transfers the solution on a coarse 
grid to a fine grid. The prolongation operator has a rank-one 
tensor structure. Let $N$-dimensional tensor 
$\tenp \in \mathbb{R}^{n_1\times n_2\times \cdots \times n_N}$
contain the function values on an $N$-dimensional tensor grid with 
$n_k$, $k = 1,2,\ldots ,N$, grid points along each direction.
The prolongation operator $\mathbf{P}^{[k]}$ to the $k$-th 
dimension is then defined as
\begin{equation} 
\label{eq: prolongation operator}
\mathbf{P}^{[k]} =
I \otimes \cdots \otimes I \otimes \underset{\text{$k$-th mode}}
{\underbrace{P^{2n_k}_{n_k}}} \otimes I \otimes \cdots \otimes I, 
\end{equation}
where $P^{2n_k}_{n_k} \in \mathbb{R}^{2n_k \times n_k}$ is the 
one-dimensional interpolation matrix defined by 
\begin{equation*}
P^{2n_k}_{n_k}= \frac{1}{2} \left(\begin{array}{cccc}
		2 &  &  & \\
		1 & 1 &  & \\
		  & 2 &  & \\
		  & 1 & 1 & \\
		  &   & \ddots & 
\end{array}\right).
\end{equation*}
If tensor $\tenp$ has the rank-$R$ separated representation 
as \eqref{eq: solution: discrete separated expansion} with 
$n= n_1 = n_2 = \cdots = n_N$, the complexity to interpolate 
a single dimension is $\mathcal{O}(n)$, and the total 
complexity of a full interpolation over $N$-dimensional tensor grid
is $\mathcal{O}(nN)$. We summarise the multi-level accelerated 
adaptive inverse power method as follows:
\begin{itemize}
\item[]
\begin{itemize}
\item[{\bf Step 0.}] 
Initialize: initial grid size on the coarsest grid $h_{1}$;  
initial guess $\tenp^{(1)}_{0}$; initial error tolerance $\varepsilon^{(1)}$; 
maximum number of grid levels $L_{max}$; and let $\ell=1$.
\item[{\bf Step 1.}] 
Solve the eigenvalue problem 
\eqref{eq: stationary equation in tensor format} on the $\ell$-th level
with initial guess $\tenp^{(\ell)}_0$, using the adaptive inverse power
method. Return the solution $\tenp^{(\ell)}_k$ that satisfies the error
tolerance 
$\|  \tenp^{(\ell)}_k -\tenp^{(\ell)}_{k-1} \| \leq \varepsilon^{(\ell)}$.
\item[{\bf Step 2.}] 
If $\ell < L_{max}$, interpolate the solution $\tenp^{(\ell)}_k$ to a finer grid
by successive application of the prolongation operator $\mathbf{P}^{[k]}$ 
in \eqref{eq: prolongation operator} to each dimension. Let 
$\tenp^{(\ell+1)}_0 = \tenp^{(\ell)}_k$, $\varepsilon^{(\ell+1)}
= \varepsilon^{(\ell)}/2$, $\ell = \ell+1$. Go to Step 1.
\item[{\bf Step 3.}] 
If $\ell = L_{max}$, return the solution $\tenp^{(\ell)}_k$ and exit.
\end{itemize}
\end{itemize}
Multilevel approach is used in Section \ref{sec20D} to analyse the
20-dimensional chemical system \eqref{eq:20D}. CPU times for each
grid size are shown in Table S\ref{tab20D}.
In general, the operators of both the CME and CFPE are 
non-symmetric and ill-conditioned, and challenging to handle using 
tensor-structured solvers~\cite{oseledets2009breaking,oseledets2010approximation}.
Although it can be improved by shortening the time-step~[48], the CFPE has 
a distinctive advantage over the CME for its flexibility in choosing the grid size, 
which enables to control the accuracy and use of acceleration strategies, 
such as the presented multilevel approach.

\paragraph*{Implementation.}
The TPA, implemented in MATLAB,
is included in the Stochastic Bifurcation Analyzer toolbox 
available at \texttt{http://www.stobifan.org}. 
The source code relies on the Tensor Train 
Toolbox~[22]. 
Simulations are performed on a 64-bit Linux desktop equipped with 
Quad-Core AMD Opteron$^{\hbox{\scriptsize \rm TM}}$ Processor 8356 
$\times$ 16 and 63 GB RAM.

\subsection{Elementary tensor operations}
\label{subsectensoroperations}

The computation of the tensor-structured parametric solution $\tenp$
has been described in Section~\ref{subsecsolveCFPE}.
In this section, we discuss computational details of
post-processing the solution in the 
form~\eqref{eq: solution: discrete separated expansion} for parametric analysis. 
This analysis is based on high-dimensional integration, 
implemented using the $k$-mode product described below.

\paragraph{Tensor multiplication: the $k$-mode product~\cite{kolda2009tensor}.} 
Let $\tenp \in \mathbb{R}^{n_1 \times n_2 \times \cdots \times n_N}$
be an $N$-dimensional tensor, the $k$-mode product of 
$\tenp$ with a vector $\mathbf{q} \in \mathbb{R}^{n_k}$
is denoted by $\tenp \times_k \mathbf{q}$ and is a tensor of
size $n_1 \times n_{k-1} \times 1 \times n_{k+1}  \times \cdots \times n_N$. 
Elementwise, we have
\begin{equation} 
\label{eq: mode product}
(\tenp \times_k \mathbf{q})
_{i_1,\ldots,i_{k-1},1,i_{k+1}, \ldots, i_N} 
= \sum_{j_k = 1}^{n_k} \tenp
_{i_1,\ldots,i_{k-1},j_k,i_{k+1}, \ldots, i_N} \, q_{j_k}.
\end{equation}
Further, if $\tenp$ can be written as
a rank-$R$ tensor, i.e.,
$
\displaystyle 
\tenp =
\sum_{r=1}^R 
\mathbf{\phi}_1^{[r]} \otimes 
\cdots 
\otimes \mathbf{\phi}^{[r]}_N,
$
then the $k$-mode product can be evaluated through 
$R$ one-dimensional inner products:
$$
\tenp \times_k \mathbf{q}
= \sum_{r=1}^R 
\mathbf{\phi}_1^{[r]} \otimes 
\cdots 
\otimes \mathbf{\phi}_{k-1}^{[r]} 
\otimes \langle \mathbf{\phi}_k^{[r]}, \mathbf{q} \rangle 
\otimes \mathbf{\phi}_{k+1}^{[r]} 
\otimes 
\cdots 
\otimes \mathbf{\phi}_N^{[r]}.
$$

\paragraph{The $(i_1,i_2,\ldots,i_N)$-th order moment computation
in equation (5).} 
For tensor-structured parametric solution in 
\eqref{eq: solution: discrete separated expansion}, integral
(5) can be simultaneously approximated for all parameter sets through 
successive application of
the mode product introduced in \eqref{eq: mode product} as
\begin{equation} 
\label{eq: moment computation in tensor format}
\mu_{[i_1,\ldots, i_N]}(\mathbf{k}^*) 
\approx
h^x_1 h^x_2 \cdots h^x_N
\left( \tenp
\times_1 \mathbf{x}_1^{i_1} 
\times_2 \mathbf{x}_{2}^{i_2}
\times_3
\cdots 
\times_N \mathbf{x}_N^{i_N}
\right),
\end{equation}
where 
$\mathbf{x}_d^{i_d} 
= 
(x_{d,1}^{i_d},
x_{d,1}^{i_d},
\dots,
x_{d,n_d}^{i_d})^T$ 
and $h^x_d$ is the grid size, defined in Section~\ref{sectens}.
The computational complexity of 
\eqref{eq: moment computation in tensor format} is $\mathcal{O}(nNR)$, 
where $n = \max\{n_1,n_2,\dots,n_d\}$.

\paragraph{Computation of integral in equation (8) in the paper.}
Given the distributions of parameters $q_j(k_j)$ for $j=1,\ldots,K$, 
the integral in equation (8) can be efficiently computed using
the tensor-structured solution
\eqref{eq: solution: discrete separated expansion} by
\begin{equation}
\label{eq:invprobintvar}
\tenp_{\mathbf{x}} =
h^k_1 h^k_2 \dots h^k_K \left( \tenp
\times_{N+1} \mathbf{q}_{1}
\times_{N+2} \mathbf{q}_{2}
\times_{N_3}
\cdots 
\times_{N+K} \mathbf{q}_{K} \right),
\end{equation}
where the entries of vectors $\mathbf{q}_{j}$ for $j=1,\ldots,K,$
represent the values of $q_j(k_j)$ at the discrete node points
$k_{j,1},$ $k_{j,2},$ $\dots,$ $k_{j,m_j}$ and $h^k_1,$ $h^k_2,$
$\dots,$ $h^k_K$ are grid sizes
in the parameter space, defined in Section~\ref{sectens}.
If $m = m_1 = m_2 = \cdots = m_K = m$ then the complexity
of evaluating the approximation \eqref{eq:invprobintvar} of 
the $K$-dimensional 
integral (8) is $\mathcal{O}(mKR)$, which scales linearly with the
separation rank $R$, the number of parameters $K$, and the number
of grid nodes $m$ along each dimension in the parameter space.

\paragraph{Computing transition probability and oscillation amplitude.}

In the parameter estimation (Figure~2) and  sensitivity analysis
(Figure~S\ref{figureS1}), we illustrate the results based on 
the transition probability and oscillation amplitude that are 
extracted from tensor-structured parametric solution. For example, 
the probability that, in steady state distribution, the 
$\ell$-th chemical species stays below a certain threshold $\tilde{x}_\ell$,
is estimated as follows. We first integrate out all the other dimensions 
in the state space, and integrate the $\ell$-th dimension up to $\tilde{x}_\ell$, 
i.e.,
$$
p(x_\ell \leq \tilde{x}_\ell \, | \, \mathbf{k}) 
=
\int_{a^x_1}^{b^x_1} \cdots 
\int_{a^x_{\ell-1}}^{b^x_{\ell-1}}
\int_{a^x_\ell}^{\tilde{x}_\ell}
\int_{a^x_{\ell+1}}^{b^x_{\ell+1}} 
\cdots 
\int_{a^x_N}^{b^x_N}
p(\mathbf{x} \, | \, \mathbf{k})\, \mathrm{d} \mathbf{x}.
$$
In tensor structure, we use $N$-mode products to 
compute $p(x_\ell \leq \tilde{x}_\ell \, | \, \mathbf{k})$ simultaneously 
for all parameter combinations by
$$
h^x_1 h^x_2 \cdots h^x_N
\left( \tenp
\times_1 \mathbf{1} 
\times_2 \mathbf{1} 
\times_3 
\cdots 
\times_{\ell-1} \mathbf{1} 
\times_\ell \mathbf{1}_{\tilde{x}_\ell}
\times_{\ell+1} \mathbf{1} 
\times_{\ell+2}
\cdots 
\times_N \mathbf{1} \right),
$$
where $\mathbf{1}$ denotes a vector of
all ones. Entries of $\mathbf{1}_{\tilde{x}_\ell}$ are equal to 1 if the
corresponding grid point is smaller or equal to $\tilde{x}_\ell$, 
while its other entries are zero.

\subsection{Sensitivity analysis}
\label{subsecsensitivity}

The sensitivity indicator for an observable quantity $\Theta$
with respect to a parameter $k$ is often computed as a finite 
difference~\cite{savageau1971parameter}
\begin{equation}
\label{eq: sensitivity finite difference}
S(\Theta) \approx \frac{ \| \Theta(k+\Delta k) - \Theta(k) \|}{\Delta k}
\frac{k}{\| \Theta(k) \|},
\end{equation}
where $\| \cdot \|$ represent a suitable norm, and $\Delta k$ 
is a change in the value of $k$.
The model is sensitive to the parameter, if a small $\Delta k$ 
yields a large value of $S(\Theta)$.
For deterministic models, the observable $\Theta$ is usually 
the steady-state mean concentration.
In stochastic setting, we have more options.

For example, let us consider the cell cycle model described in
Figure~4(a). We will study the sensitivity with respect to the parameter 
$k_1$ for the following three observables $\Theta$: mean concentration 
of the MPF ($\Theta_m$), the oscillation amplitude ($\Theta_a$) and 
the steady state distribution ($\Theta_p$).
In the case of the oscillation amplitude, we quantify $\Theta_a$ 
as the probability that the molecular population of the active MPF 
exceeds 400.

In the TPA framework, $\Theta_m$ and $\Theta_a$ are evaluated for all 
considered values of $k_1$ with computational cost scaling linearly 
with $N$.
More importantly, the tensor-structured data enable direct comparison 
of two steady state probabilities in the 6-dimensional state space. 
Namely, the norm $\| \Theta_p(k_1 + \Delta k_1) - \Theta_p(k_1) \|$, 
needed in \eqref{eq: sensitivity finite difference}, can be directly computed.
The results are plotted in Figure~S\ref{figureS1}.
We observe that, within the considered range of $k_1$ 
(see Table~S\ref{tabcellcycleranges}), the sensitivity in the steady 
state distribution (blue curve) dominates in magnitude over 
$S(\Theta_m)$ and $S(\Theta_a)$. The steady state distribution
contains a global information about the system and is more
sensitive to parameter changes than derived quantities, 
like $\Theta_m$ and $\Theta_a$. 

\section{Description of models used in the illustrative TPA computations}
\label{secbiomodels}

\subsection{Schl\"{o}gl model}

The Schl\"{o}gl system is defined by chemical reactions listed in 
Table~S\ref{tableschlpar}. This table also shows the true values 
of parameters $k_1$, $k_2$, $k_3$ and $k_4$. Using this system, 
we illustrate the capabilities of the TPA for the parameter 
estimation and identifiablity. Table~S\ref{tableschlmom} provides
the values of the first three statistical moments and the corresponding 
weights. The moments have been computed from a time series obtained 
by a long-time stochastic simulation with the values of parameters
given in Table~S\ref{tableschlpar}. A short segment of 
the time series is illustrated in Figure~1(a).
We then assume that the values of parameters $k_1$, $k_2$, $k_3$
and $k_4$ are unknown. The TPA enables to evaluate the moment matching 
distance function given in the paper in equation (3), for all values of 
parameters within the parameter space given in Table~S\ref{tabschlogl}. 
The resulting data are stored in the tensor format which enables 
efficient manipulations and post-processing.

Having the values of the distance function stored in the tensor format,
we can easily and quickly obtain further pieces of information.
For example, we can find those parameter values which do not fit 
the moments exactly, but with certain accuracy. More precisely, 
we consider tolerance $J_\mathrm{TOL} = 0.25\,\%$ and visualize 
in Figures~2(a)--(d) parameter values with distance
function $J$ less than $J_\mathrm{TOL}$.
Alternatively, if the values of moments are not available, we 
can utilize the TPA for different experimental data -- see Figure~1(b)
and equation (6) in the paper. 

{\modify
Note that all four values of parameters $k_1,$ $k_2$, $k_3$ and
$k_4$ cannot be estimated
solely from the steady state distribution, because it 
does not inform us how fast the 
system reaches the steady state. In particular, 
if $\{k_i\}_{i = 1,2,3,4}$ fit the pseudo-experimental data, 
then $\{Ck_i\}_{i = 1,2,3,4}$ for any $C > 0$ fit these data as well. 
Therefore, in Figures 1(b) and 2, 
we fix one of the parameters at its true value 
and estimate values of the other three.
}

\begin{table}[h!]
\begin{center}
\caption{{\it Overview of kinetic reactions of the Schl\"{o}gl model.}
\label{tableschlpar}}
\begin{tabular}{cccl}
\hline
Index & Reaction & Kinetic rate$^a$ & True value \\
\hline
1 & $3X \rightarrow 2X$ & $k_1/V^2$ & $k_1 = 2.5 \times 10^{-4}$ \\
2 & $2X \rightarrow 3X$ & $k_2/V$ & $k_2 = 0.18$ \\
3 & $\emptyset \rightarrow X$ & $k_3\times V$ & $k_3 = 2250$ \\
4 & $X \rightarrow \emptyset$ & $k_4$ & $k_4 = 37.5$ \\
\hline		
\end{tabular}
\\$^a$The reacting volume is set to $V=1$ unit. 
\end{center}
\end{table}

\begin{table}[h!]
\begin{center}
\caption{{\it Moments estimated from stochastic simulation
of the Schl\"{o}gl model.}
\label{tableschlmom}}
\begin{tabular}{cll}
\hline
Moment order & Value & Weight \\ \hline
1      & $\hat\mu_1 = 261.32$             &  $\beta_1 = 1$  \\
2      & $\hat\mu_2 = 2.03 \times 10^4$   &  $\beta_2 = 100$  \\
3      & $\hat\mu_3 = -2.04 \times 10^5$  &  $\beta_3 = 0.001$   \\  \hline
\end{tabular}
\end{center}
\end{table}
		
\begin{table}[h!]
\begin{center}
\caption{\label{tabschlogl}
{\it Properties of molecular and rate variables in the Schl\"{o}gl model.}}
\begin{tabular}{cccc}
\hline
Type & Notation & Range & No. of nodes \\
\hline
Species & $X$ & $[0, 1000]$ & 1024 \\
Rate & $k_1$ & $[2.43 \times 10^{-4}, 2.58 \times 10^{-4}]$ & 128 \\
Rate & $k_2$ & $[0.17,0.19]$ & 128 \\
Rate & $k_3$ & $[2134,2266]$ & 128 \\
Rate & $k_4$ & $[36.08,38.63]$ & 128 \\
\hline		
\end{tabular}
\end{center}
\end{table}

\subsection{Cell cycle model}

\begin{figure}[t]
\begin{center}
\includegraphics[width=.5\textwidth]{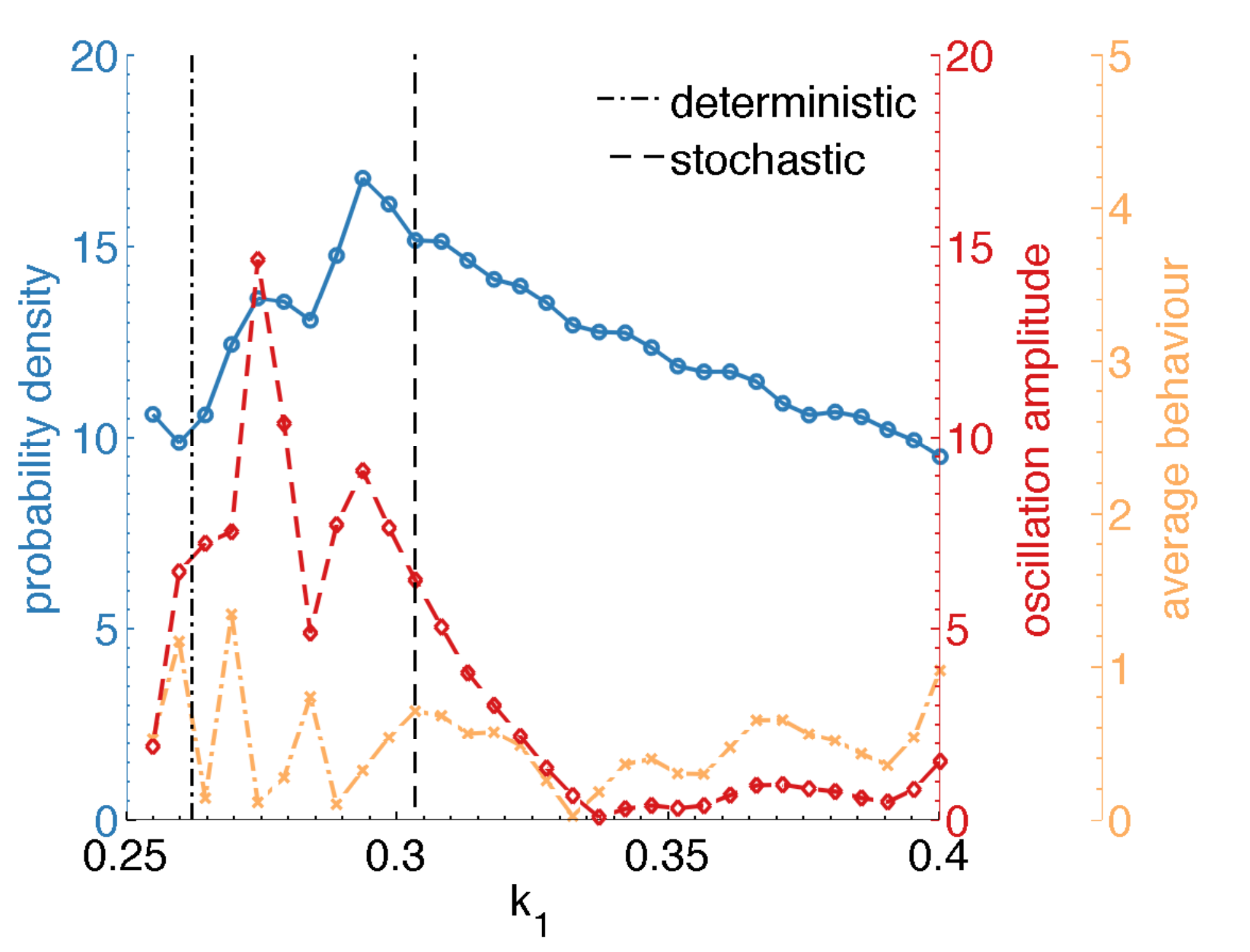}
\end{center}
\caption{
{\it Sensitivity indicators $S(\Theta)$ calculated by
$\eqref{eq: sensitivity finite difference}$ for $32$ equidistant nodes 
within the range $[0.25, 0.4]$ of parameter $k_1$ 
(see Table~S$\ref{tabcellcycleranges}$ and Section~$\ref{subsecsensitivity}$). 
Three observables are considered: stationary distribution $\Theta_p$
(blue), oscillation amplitude $\Theta_a$ (red) and average number 
$\Theta_m$ of active MPF (orange).
The dot-dashed and dashed lines indicate parameter values for 
which Figures}~4(b) and 4(c) {\it in the paper were computed,
i.e. $k_1 =0.2694$ (deterministic bifurcation point) 
and $k_1 = 0.3032$, respectively.}
\label{figureS1}
}
\end{figure}

The cell cycle model consists of nine chemical reactions and six 
chemical species
as listed in Tables~S\ref{tabcellcycle} and S\ref{tabcellcycleranges},
see also Figure~4(a) in the paper. We have used this model to show 
how the TPA can be used to analyse bifurcations for high-dimensional 
problems, see Figures~4(b), 4(c) and 5 in the paper.
In addition, we have used this system to discuss the sensitivity of various
quantities in the stochastic model, see Figure~S\ref{figureS1} 
and Section~\ref{subsecsensitivity}.

\begin{table}[h!]
\begin{center}
\caption{\label{tabcellcycle}
{\it Overview of kinetic reactions of the cell cycle model.}}
\begin{tabular}{cccl}
\hline
Index & Reaction & Kinetic rate$^a$ & Parameter(s) \\ \hline
1 &  $M \rightarrow C_2 + YP$ & $k_1$  & bifurcation parameter \\
2 &  $\emptyset \rightarrow Y$ & $k_2\times V$ &  $k_2 = 0.015$ \\
3 &  $CP + Y \rightarrow pM$ & $k_3/V$ & $k_3 = 200$  \\
4 & $pM \rightarrow M$ & $k_4' + k_4 \left( M/V \right)^2$ & $k_4 = 180$,  		$k_4' = 0.018$   \\
5 &  $M \rightarrow pM$ & $k_5 \times tP$ & $k_5 = 0$, $tP=0.001$ \\
6 &  $Y \rightarrow \emptyset$  & $k_6$  &  $k_6 = 0$ \\
7 &  $YP \rightarrow \emptyset$ & $k_7$ & $k_7 = 0.6$ \\
8 &  $C_2 \rightarrow CP$ & $k_8 \times tP$  & $k_8 = 1000$ \\
9 &  $CP \rightarrow C_2$ & $k_9$ & $k_9 = 1000$ \\ \hline
\end{tabular} 
\\$^a$The volume corresponds to a single
cell and is set to $V = 5000$ units.
\end{center}
\end{table}

\begin{table}[h!]
\begin{center}
\caption{\label{tabcellcycleranges}
{\it Properties of molecular and rate variables in the cell cycle model.}}
\begin{tabular}{ccccc}
\hline
Type & Name & Notation & Range & No. of nodes \\ \hline
Species & cdc2 & C2 & $[2230, 4990]$ & N/A$^a$ \\
Species & cdc2-P & CP & $[10, 70]$ & 256 \\
Species & p-cyclin-cdc2-p & pM & $[0, 1500]$ & 256 \\
Species & p-cyclin-cdc2 & M & $[0, 1200]$ & 256 \\
Species & cyclin & Y & $[20, 70]$ & 256 \\
Species & p-cyclin & YP & $[0, 700]$ & 256 \\
Rate & degradation rate of active MPF  & $k_1$ & $[0.25, 0.4]$ & 64 \\  \hline
\end{tabular}
\\$^a$Discretisation of cdc2 is not applicable here, 
since this variable is eliminated by the conservation law 
of cdc2 assumed by the original author.
\end{center}
\end{table}

\subsection{FitzHugh-Nagumo model}

The FitzHugh-Nagumo model consists of five chemical reactions between 
two chemical species. It is illustrated in Figure~6(a) in the paper and 
the parameter ranges and mean values are provided in Tables~S\ref{FHNrange} 
and S\ref{FHNparam}. This model is used to show how
the TPA can assess the influence of the extrinsic noise,
see Figure~6 in the paper. 

\begin{table}[h!]
\begin{center}
\caption{{\it Properties of molecular and rate variables in the 
FitzHugh-Nagumo model.}
\label{FHNrange}}
\begin{tabular}{cccc}
\hline
Type & Notation & Range & No. of nodes \\
\hline
Species & $X_1$ & $[0, 1800]$ & 256 \\
Species & $X_2$ & $[0, 700]$ & 256 \\
Rate & $k_1$ & $[0.17, 0.23]$ & 128 \\
Rate & $k_2$ & $[0.952, 0.1288]$ & 128 \\
Rate & $k_3$ & $[2.125, 2.875]$ & 128 \\
Rate & $k_4$ & $[0.0892, 0.1207]$ & 128 \\
\hline		
\end{tabular}
\end{center}
\end{table}

\begin{table}[h!]
\begin{center}
\caption{{\it Overview of kinetic reactions of the FitzHugh-Nagumo model.}
\label{FHNparam}}
\begin{tabular}{cccl}
\hline
Index & Reaction & Kinetic rate$^a$ & Mean value \\
\hline
1 & $X_1 \rightarrow 2X_1$ & $(X_1 - k_1\times V)(V - X_1)$ & $k_1 = 0.2$ \\
2 & $X_1 \rightarrow \emptyset$ & $X_2$ & N/A \\
3 & $\emptyset \rightarrow X_1$ & $k_2\times V$ & $k_2 = 0.112$ \\
4 & $X_2 \rightarrow \emptyset$ & $k_3 \times k_4$ & $k_3 = 2.5$\\
5 & $X_1 \rightarrow X_1 + X_2$ & $k_4$ & $k_4 = 0.105$ \\
\hline		
\end{tabular}
\\$^a$The system volume is $V = 2000$ units.
\end{center}
\end{table}

\subsection{A chemical reaction system in 20 dimensions}
\label{sec20D}

We consider a reaction chain of 20 molecular species:
\begin{equation}
\label{eq:20D}
\mbox{ \raise 0.851 mm \hbox{$\emptyset$}}
\;
\mbox{ \raise 0.851 mm \hbox{$\displaystyle \overset{k_{0}}{\longrightarrow}$}}
\;
\mbox{ \raise 0.851 mm \hbox{$X_1$}}
\;
\mathop{\stackrel{\displaystyle\longrightarrow}\longleftarrow}^{k_1}_{k_{-1}}
\;
\mbox{\raise 0.851 mm\hbox{$X_2$}}
\;
\mathop{\stackrel{\displaystyle\longrightarrow}\longleftarrow}^{k_2}_{k_{-2}}
\;
\mbox{\raise 0.851 mm\hbox{$\cdots$}}
\;
\mathop{\stackrel{\displaystyle\longrightarrow}\longleftarrow}^{k_{18}}_{k_{-18}}
\;
\mbox{\raise 0.851 mm\hbox{$X_{19}$}}
\;
\mathop{\stackrel{\displaystyle\longrightarrow}\longleftarrow}^{k_{19}}_{k_{-19}}
\;
\mbox{\raise 0.851 mm\hbox{$X_{20}$}}
\;
\mbox{ \raise 0.851 mm \hbox{$\displaystyle \overset{k_{20}}{\longrightarrow}$}}
\;
\mbox{ \raise 0.851 mm \hbox{$\emptyset$,}}
\end{equation}
where $k_0=12$, $k_i=0.2$ for $i=1, \ldots, 20$,
and $k_{-j}=0.1$ for $j=1, \ldots, 19$.
A multilevel approach is implemented to solve the underlying CFPE, 
where the steady state distribution is first approximated on a 
coarse grid, and then interpolated to a finer grid 
(see Section~\ref{subsecsolveCFPE}). The results are plotted 
in Figure 7. 

\begin{table}[ht]
\begin{center}
\caption{\label{tab20D}
{\it Multilevel discretisation for the $20$-dimensional reaction chain}
\eqref{eq:20D}.}
\begin{tabular}{cccccccc}
\hline
Level               & 1    & 2     & 3    & 4     & 5       & 6        
& 7 \\ \hline
No. of nodes $n$  &  8   & 16   & 32  & 64   & 128  & 256    & 512\\
Grid size $h$        
& 20  & 10   & 5    & 2.5  & 1.25  & 0.625 & 0.3125 \\
CPU time ($\times10^3$ sec) & 2.82 & 7.02 & 2.84 & 2.3& 2.25 & 0.08 & 0.10\\
\hline
\end{tabular} 
\end{center}
\end{table}

\end{document}